\documentclass{aastex}
\usepackage{emulateapj5,times,mathptm}

\newcommand {\asec} {$^{\prime\prime}$}

\def\amin{\ifmmode ^{\prime}\else$^{\prime}$\fi}
\def\asec{\ifmmode ^{\prime\prime}\else$^{\prime\prime}$\fi}

\def\etal{{et\,al.\,}}

\def\asca{{\it ASCA\/}}

\def\chandra{{\it Chandra\/}}

\def\heao1{{\it HEAO-1\/}}

\def\rosat{{\it ROSAT\/}}

\def\ltsima{$\; \buildrel < \over \sim \;$}
\def\simlt{\lower.5ex\hbox{\ltsima}}
\def\gtsima{$\; \buildrel > \over \sim \;$}
\def\simgt{\lower.5ex\hbox{\gtsima}}

\begin{document}
%%%%%%%%%%%%%%%%%%%%%%%%%%%%%%%%%%%%%%%%%%%%%%%%%%%%%%%%%%%%%%%%%%%%%%%%%%%%%%%%%%
%

%
%%%%%%%%%%%%%%%%%%%%%%%%%%%%%%%%%%%%%%%%%%%%%%%%%%%%%%%%%%%%%%%%%%%%%%%%%%%%%%%%%%
\title{The Chandra Deep Field North Survey. X. X-ray Emission from Very Red Objects}
%%%%%%%%%%%%%%%%%%%%%%%%%%%%%%%%%%%%%%%%%%%%%%%%%%%%%%%%%%%%%%%%%%%%%%%%%%%%%%%%%%
%

\author{D.M.~Alexander, C.~Vignali, F.E.~Bauer, W.N.~Brandt, A.E.~Hornschemeier, G.P.~Garmire and D.P.~Schneider}

\affil{Department of Astronomy \& Astrophysics, 525 Davey Laboratory, The Pennsylvania State University, University Park, PA 16802}

%\vfill\eject

\shorttitle{X-ray emission from VROs in the Chandra Deep Field North Survey}

\shortauthors{Alexander et al.}

%
%%%%%%%%%%%%%%%%%%%%%%%%%%%%%%%%%%%%%%%%%%%%%%%%%%%%%%%%%%%%%%%%%%%%%%
\begin{abstract}
%%%%%%%%%%%%%%%%%%%%%%%%%%%%%%%%%%%%%%%%%%%%%%%%%%%%%%%%%%%%%%%%%%%%%%
%

The multi-wavelength properties of Very Red Objects (VROs; $I-K\ge4$) are largely unknown as many of these sources are optically faint ($I\ge24$) and undetected at most wavelengths. Here we provide constraints on the X-ray (0.5--8.0~keV) properties of VROs using the 1~Ms \chandra\ exposure of an $8.4\amin\times8.4\amin$ region within the Hawaii Flanking-Field area containing the Hubble Deep Field North (HDF-N). We find that VROs detected in the hard band (2.0--8.0~keV) have flat X-ray spectral slopes ($\Gamma\approx$~0.9) and X-ray properties consistent with those expected from luminous obscured AGN. The fraction of such sources in the $K\le20.1$ VRO population is $14^{+11}_{-7}$\%. Conversely, the average X-ray spectral slope of VROs detected in the soft band (0.5--2.0~keV) but not in the hard band is comparatively steep ($\Gamma>1.4$), and the X-ray emission from these sources is consistent with that expected from less energetic processes (i.e.,\ star formation, low-luminosity AGN activity, normal elliptical galaxy emission); star-formation and low-luminosity AGN activity scenarios are favored in those sources with irregular optical morphologies. Stacking analyses of the X-ray emission from VROs not individually detected at X-ray energies yield significant detections ($\ge$99\% confidence level) in the soft band and the full band (0.5--8.0~keV). We find this X-ray emission is produced predominantly by the optically brightest VROs. The simplest explanation of this result is that we have detected the average X-ray emission from non-active VROs with low X-ray-to-optical flux ratios [$\log{({{f_{\rm X}}\over{f_{\rm I}}})}\approx$~--2]; this is consistent with that expected if the majority of these VROs are $\approx{M^{*}_{I}}$ elliptical galaxies. A number of VROs are also detected with mid-infrared (15~$\mu$m) and radio emission, and we provide constraints on the nature of this emission.

\end{abstract}

\keywords{cosmology: observations --- galaxies: active --- galaxies: starburst --- surveys --- X-rays: galaxies}

%
%%%%%%%%%%%%%%%%%%%%%%%%%%%%%%%%%%%%%%%%%%%%%%%%%%%%%%%%%%%%%%%%%%%%%%
\section{Introduction}\label{intro}
%%%%%%%%%%%%%%%%%%%%%%%%%%%%%%%%%%%%%%%%%%%%%%%%%%%%%%%%%%%%%%%%%%%%%%
%

Understanding the properties and nature of Very Red Objects (hereafter VROs\footnote{In many studies these sources are also referred to as Extremely Red Objects (EROs).}) remains a challenge to observational cosmology. This is principally because VROs are often optically faint (i.e.,\ $I\ge24$) and undetected at most wavelengths to faint limits. Although the definition of a VRO varies from study to study, it is generally accepted that a source with $R-K\ge5.3$ or $I-K\ge4$ is a VRO. These colors are consistent with those of a passively evolving elliptical galaxy at $z\simgt$1 (e.g.,\ see Pozzetti \& Mannucci 2000; Daddi \etal 2000). Optical and near-IR spectroscopy and morphology studies have indeed shown that $z\simgt$1 elliptical galaxies appear to comprise at least 40--50\% of the VRO population (e.g.,\ Cimatti \etal 1999; Moriondo \etal 2000; Mannucci \etal 2001). However, sub-millimeter and infrared observations have also demonstrated that the VRO population contains dust-enshrouded galaxies at $z\simgt$1 (e.g.,\ Cimatti \etal 1998; Barvainis, Antonucci, \& Helou 1999; Dey \etal 1999; Smail \etal 1999; Afonso \etal 2001; Elbaz \etal 2001; Pierre \etal 2001; Smith \etal 2001a). These dusty VROs may dominate the extreme red-end of the VRO population (i.e.,\ $I-K\ge5$; Smail \etal 1999; Smith \etal 2001b), although the current statistics are limited. Due to the extreme faintness of VROs, the earliest studies of source densities were limited to small areas and susceptible to ``cosmic variance'' (e.g.,\ see Table~1 in Scodeggio \& Silva 2000). Indeed, recent wide-area VRO surveys have shown evidence for source clustering (e.g.,\ Daddi \etal 2000, 2001).

Deep \rosat\ observations first showed that VROs are detectable at X-ray energies (e.g.,\ Newsam \etal 1997; Lehmann \etal 2000, 2001). Deeper X-ray surveys performed with the {\it Chandra X-ray Observatory} (hereafter \chandra; Weisskopf \etal 2000) have uncovered more examples of X-ray detected VROs (e.g.,\ Alexander \etal 2001, hereafter Paper VI, Barger \etal 2001; Brandt \etal 2001b, hereafter Paper IV; Cowie \etal 2001; Crawford \etal 2001; Hornschemeier \etal 2001, hereafter Paper II; Stanford \etal 2001). In the majority of cases this X-ray emission is consistent with that expected from obscured AGN activity. In Paper II, a 221.7~ks observation of the \chandra\ Deep Field North (hereafter CDF-N) was used to investigate the X-ray emission from a sample of $K_{\rm s}$-band selected VROs. Four of the 33 sources were individually detected at X-ray energies although only in the hard band (2.0--8.0~keV), suggesting obscured AGN activity. Conversely, stacking analyses of the X-ray undetected VROs produced a significant average detection in only the soft band (0.5--2.0~keV). The properties of this X-ray emission were found to be consistent with those expected from normal elliptical galaxies at $z\approx$~1 (i.e.,\ X-ray binary and thermal gas emission), suggesting a dichotomy in the X-ray properties of VROs. However, the definition of a VRO in Paper II was ${\cal R}-K_{\rm s}\ge5.0$, and none of the individually detected X-ray sources had ${\cal R}-K_{\rm s}\ge5.3$.

In this study we provide specific constraints on the X-ray properties of VROs using the stricter VRO definition of $I-K\ge4$ (equivalent to $R-K\ge5.3$) and the much deeper 1~Ms observation of the CDF-N (e.g.,\ Brandt \etal 2001a, hereafter Paper~V). We also exploit the deep and extensive multi-wavelength observations within the HDF-N and environs to provide constraints on the multi-wavelength properties of VROs. The Galactic column density along this line of sight is $(1.6\pm 0.4)\times 10^{20}$~cm$^{-2}$ (Stark \etal 1992). We adopt the cosmology used in Paper VI (i.e.,\ $H_0=70$~km~s$^{-1}$ Mpc$^{-1}$ and $q_0=0.1$) throughout this paper. All coordinates in this paper are J2000.

%
%%%%%%%%%%%%%%%%%%%%%%%%%%%%%%%%%%%%%%%%%%%%%%%%%%%%%%%%%%%%%%%%%%%%%%
\section{VRO sample definitions and basic properties}
%%%%%%%%%%%%%%%%%%%%%%%%%%%%%%%%%%%%%%%%%%%%%%%%%%%%%%%%%%%%%%%%%%%%%%
%

In this study we have defined three VRO samples with different $K$-band (or equivalent) magnitude limits. The first sample includes all VROs with $HK^{\prime}\le20.4$ detected in the $8.4\amin\times8.4\amin$ area used in Paper VI (hereafter referred to as the reduced Hawaii Flanking-Field area) and is referred to as the {\it moderate-depth VRO} sample (see \S2.1).\footnote{As the relationship between the $K$ band and $HK^\prime$ band is $HK^\prime-K=0.13+0.05(I-K)$ (Barger \etal 1999), this is equivalent to $K<20.1$.} The second sample includes all VROs with $K\le22$ detected in the $\approx$~5.3 arcmin$^{2}$ region of the HDF-N itself and is referred to as the {\it deep VRO} sample (see \S2.2). These two samples are complete VRO samples and allow us to estimate the fraction of X-ray emitting VROs within the VRO source population. The final VRO sample is an {\it intermediate-depth X-ray detected VRO} sample that covers the same $8.4\amin\times8.4\amin$ area as the moderate-depth VRO sample but only includes X-ray detected VROs with $HK^{\prime}\le21.4$ (see \S2.3). We have defined this sample to improve the statistics of X-ray detected VROs and provide a more detailed understanding of the X-ray properties of VROs. We are unable to construct a reliable parent VRO sample at the $\approx2\sigma$ magnitude limit of this sample without introducing a significant number of spurious sources. See Table~1 for the sample definitions. Thumbnail images of the 42 VROs used in this study are shown in Figure~1.

%
%%%%%%%%%%%%%%%%%%%%%%%%%%%%%%%%%%%%%%%%%%%%%%%%%%%%%%%%%%%%%%%%%%%%%%
% 1 Thumbnail images of all the X-ray sources
%%%%%%%%%%%%%%%%%%%%%%%%%%%%%%%%%%%%%%%%%%%%%%%%%%%%%%%%%%%%%%%%%%%%%%
%

\begin{figure*}
\vspace{0.0in}
\centerline{\includegraphics[angle=0,width=20.0cm]{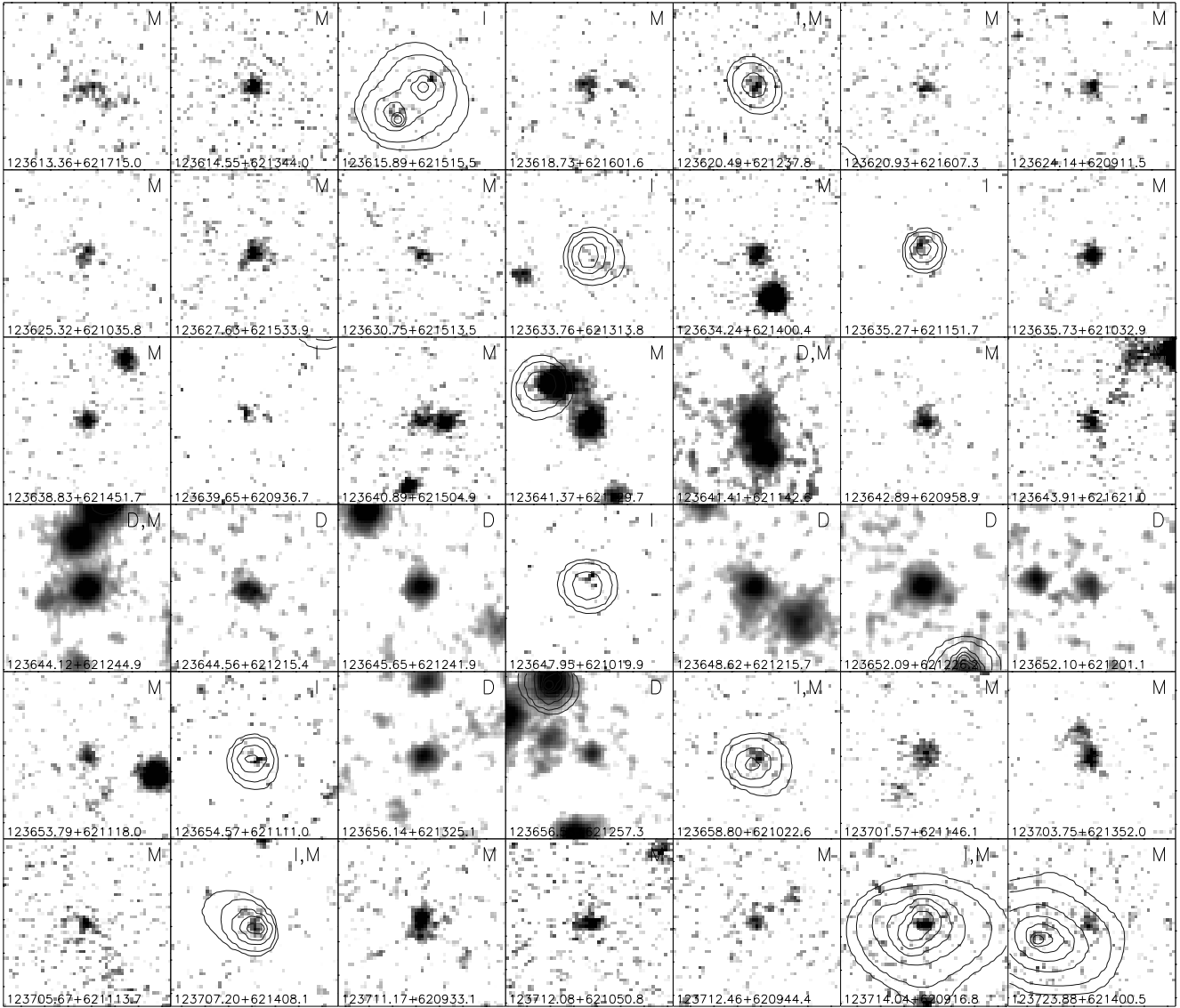}}
\vspace{0.5in}
\figcaption{Thumbnail $HK^{\prime}$-band images with overlaid X-ray contours for each VRO source; the source of interest is always at the center of the thumbnail. Each thumbnail is $9.81\asec$ on a side. The wide-field $HK^{\prime}$-band image from Barger \etal (1999) was used for the moderate-depth VRO sample and the intermediate-depth X-ray detected VRO sample, and the deep $HK^{\prime}$-band image from Barger \etal (1999) was used for the deep VRO sample. The contour levels refer to the counts detected in the full band; each X-ray image has been adaptively smoothed at the $2.5\sigma$ level using the code of Ebeling \etal (2001). The faintest X-ray sources are below the significance level of the smoothing and thus are not visible in these figures. The VRO sample from which each source is drawn is indicated in the top right-hand corner of each image: ``M'' refers to the moderate-depth VRO sample, ``I'' refers to the intermediate-depth X-ray detected VRO sample, and ``D'' refers to the deep VRO sample.}
\label{sourceexp}
\end{figure*}

%%%%%%%%%%%%%%%%%%%%%%%%%%%%%%%%%%%%%%%%%%%%%%%%%%%%%%%%%%%%%%%%%%%%%%%%%%%%%%%%%%

\subsection{The moderate-depth VRO sample}

As stated above, the area in which the moderate-depth VRO sample is defined is the reduced Hawaii Flanking-Field area used in Paper VI. However, to reduce the number of spurious VROs, we have further excluded sources within 17\asec\ of the positions of three bright $HK^\prime<14$ stars where we could not reliably identify VROs. Although optical-to-near-IR catalogs are available for this region (Barger \etal 1999)\footnote{These images are publicly available at http://www.ifa.hawaii.edu/$\sim$cowie/hdflank/hdflank.html.}, to retain consistency with our previous studies we have used the $I$-band and $HK^{\prime}$-band photometric catalogs (with $\approx$~2$\sigma$ limits of $I=25.3$ and $HK^{\prime}=21.4$, respectively) produced in Paper VI. As this region also has $V$-band coverage (Barger \etal 1999), we have produced a $V$-band catalog of sources down to $\approx$~2$\sigma$ magnitudes of $V=26.5$ using the {\sc sextractor} photometry tool (Bertin \& Arnouts 1996) and assuming the ``Best'' magnitude criteria.

To ensure a low fraction of spurious VROs we use the 5$\sigma$ magnitude limit of $HK^\prime=20.4$. We find 29 VROs with $I-HK^{\prime}\ge3.7$ (equivalent to $I-K\ge4.0$) and $HK^\prime\le20.4$ within this 70.3~arcmin$^{2}$ region, corresponding to a source density of 1500$^{+350}_{-250}$ deg$^{-2}$ (see Tables~1 and 2).\footnote{All errors are taken from Tables 1 and 2 of Gehrels (1986) and correspond to the $1\sigma$ level; these were calculated assuming Poisson statistics.} We find this VRO source density to be consistent with that found in other studies (e.g.,\ see Table~1 of Scodeggio \& Silva 2000), although the scatter from study to study is large.

\subsection{The deep VRO sample}

The deep VRO sample covers the region of the HDF-N itself. We have used the seven-band photometric catalog of Fernandez-Soto \etal (1999) to construct the deep VRO sample which has a 5$\sigma$ magnitude limit of $K=22$. The optical observations used in this catalog were taken with the {\it Hubble Space Telescope} ({\it HST}; Williams \etal 1996), and the  near-IR observations were taken with ground-based telescopes (Dickinson \etal 2000). We find nine VROs with $I-K\ge4$ and $K\le22$ within this $\approx$~5.3 arcmin$^{2}$ region (see Table~3), corresponding to a source density of 6100$^{+2800}_{-2000}$ deg$^{-2}$ (see Table~1).\footnote{We converted the {\it HST} $F814W$-band photometry to the standard $I$ band using $I=-0.39+AB_{814}-0.09(AB_{606}-AB_{814})$ (Cowie, Songaila, \& Barger 1999).} There are few constraints on the source density of VROs at these faint $K$-band magnitudes, and all studies performed to date have covered small areas (e.g.,\ see Table~5 in McCracken \etal 2000). However, our source density is in reasonable agreement with previous studies given the probable large effects of ``cosmic variance''.

Five sources within the deep VRO sample have $K$-band magnitudes and colors that fall within the moderate-depth VRO sample definition: two sources are included in both samples, and the other three sources fall just outside the VRO sample definition with the moderate-depth VRO sample photometry. These small differences in source magnitudes are to be expected given the faintness of these VROs. There are no sources within the moderate-depth VRO sample that lie within the HDF-N and are not included in the deep VRO sample; this is probably due to a statistical fluctuation as there does not appear to be any systematic differences in the photometry of these two samples.

\subsection{The intermediate-depth X-ray detected VRO sample}

The intermediate-depth X-ray detected VRO sample covers the same area as the moderate-depth VRO sample and includes all X-ray detected VROs down to the $\approx$~2$\sigma$ magnitude limit of $HK^{\prime}=21.4$. We are unable to define a parent VRO sample at this magnitude limit without introducing a significant number of spurious sources. However, due to the relatively small fraction of VROs with X-ray counterparts, we significantly reduce the probability of spurious $HK^{\prime}$ sources by selecting those sources with X-ray detections.

The X-ray data used to produce this sample were taken from the 1 Ms CDF-N catalogs (see Paper V), which contain point sources down to on-axis flux limits of $\approx3\times10^{-17}$~erg~cm$^{-2}$~s$^{-1}$ and $\approx2\times10^{-16}$~erg~cm$^{-2}$~s$^{-1}$ in the soft and hard bands, respectively. Within the reduced Hawaii Flanking-Field area that defines this sample there are 141 X-ray sources (see Paper VI). We show the adaptively smoothed full-band (\hbox{0.5--8.0~keV}) \chandra\ image of this area in Figure~2. These X-ray sources are matched to $I$-band and $HK^{\prime}$-band counterparts with a matching radius of 1\asec\ (in practice all sources had a matching distance $<0.8$\asec) to be consistent with Paper VI; we would conservatively expect 0.2 spurious matches with this criterion. In total there are 10 X-ray detected VROs with $I-K\ge4$ and $HK^{\prime}\le21.4$, corresponding to a source density of 500$^{+200}_{-150}$ deg$^{-2}$ (see Tables~1 and 4 and Figure~2). Six of these sources are exclusively in the intermediate-depth X-ray detected VRO sample (the other four sources are also within the moderate-depth VRO sample) and have magnitudes close to the $\approx$~2$\sigma$ limit of $HK^{\prime}=21.4$; we would expect 0.3 of these $HK^{\prime}$ sources to be spurious.

%
%%%%%%%%%%%%%%%%%%%%%%%%%%%%%%%%%%%%%%%%%%%%%%%%%%%%%%%%%%%%%%%%%%%%%%
% 2 Chandra image - B/W
%%%%%%%%%%%%%%%%%%%%%%%%%%%%%%%%%%%%%%%%%%%%%%%%%%%%%%%%%%%%%%%%%%%%%%
%

\begin{figure*}
\vspace{0.0in}
\centerline{\includegraphics[width=15.0cm]{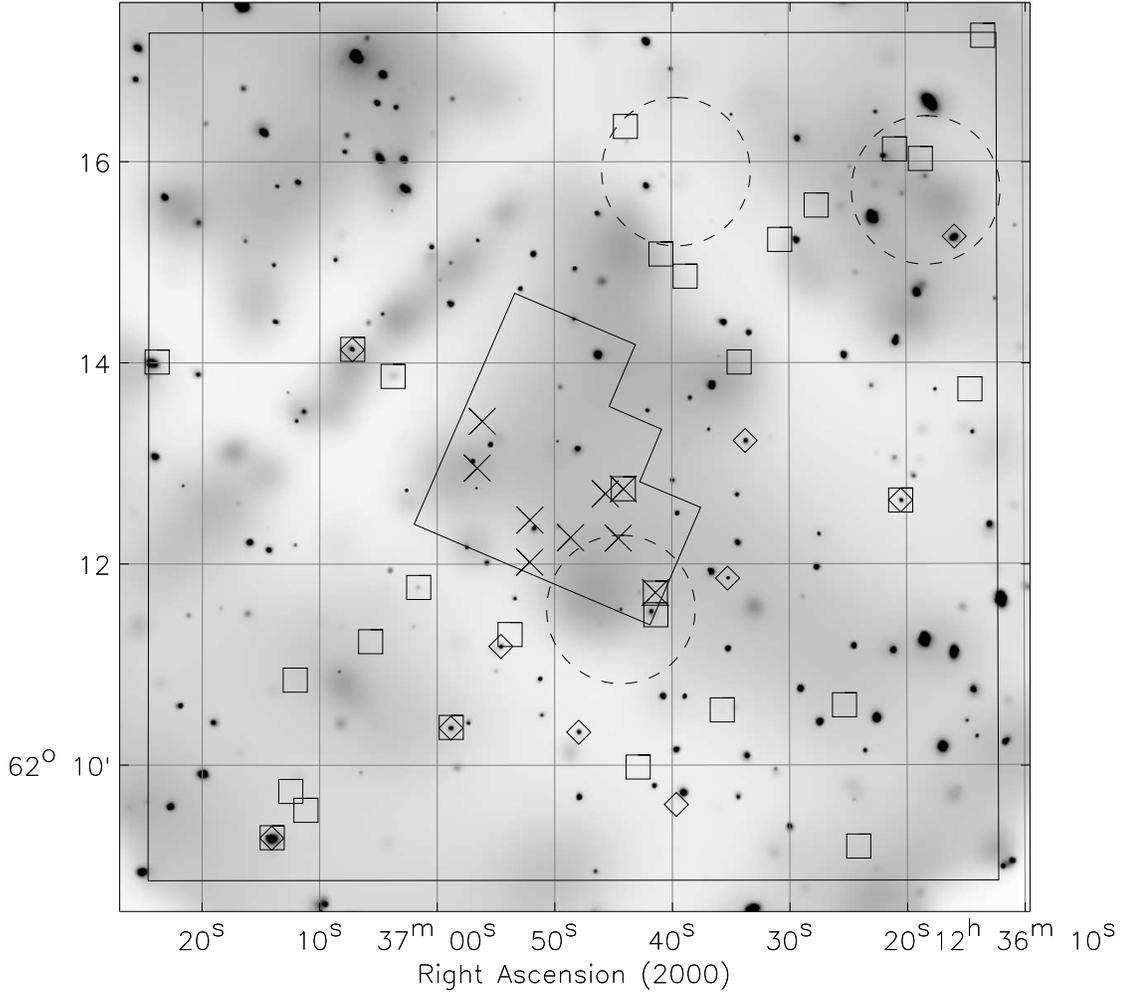}}
\vspace{0.5in}
\figcaption{Adaptively smoothed full-band 1~Ms \chandra\ image. The squares show the positions of the sources in the moderate-depth VRO sample, the diamonds show the positions of the sources in the intermediate-depth X-ray detected VRO sample, and the crosses show the positions of the sources in the deep VRO sample. This image has been made using the standard \asca\ grade set and has been adaptively smoothed at the $2.5\sigma$ level using the code of Ebeling \etal (2001). The HDF-N is shown as the polygon at the center of the image, the large box indicates the reduced Hawaii Flanking-Field area used in this study, and the dashed circles indicate the positions and adopted sizes of the cluster and cluster-candidate regions described in \S3.5. Most of the apparent diffuse emission is instrumental background. The faintest X-ray sources are below the significance level of the smoothing and thus are not visible in this figure.}
\label{sourceexp}
\end{figure*}

%%%%%%%%%%%%%%%%%%%%%%%%%%%%%%%%%%%%%%%%%%%%%%%%%%%%%%%%%%%%%%%%%%%%%%

Only one source is an extreme VRO (i.e.,\ $I-K\ge5$), although three sources are undetected in the $I$ band and have lower limits on their $I-K$ colors. The fraction of VROs within the X-ray source population at these X-ray depths is $8^{+3}_{-2}$\% and, with the exception of one source, all are optically faint with $I\ge24$ (see Paper VI).

\subsection{X-ray detected VROs in the moderate-depth and deep VRO samples}

Four of the 29 VROs in the moderate-depth VRO sample are detected at X-ray energies, and none of the nine VROs in the deep sample is detected at X-ray energies (see Tables~1--4). All these X-ray sources were detected by {\sc wavdetect} (Freeman \etal 2002) with a false-positive probability threshold of $10^{-7}$ in at least one X-ray band (Paper V).

However, given the comparitively low source density of VROs, we are able to search for lower significance X-ray sources associated with VROs without introducing a significant number of spurious X-ray sources. We ran {\sc wavdetect} with a false-positive probability threshold of 10$^{-4}$ and found two further X-ray sources with positions within 1\asec\ of sources in the moderate-depth VRO sample as well as one X-ray source with a position within 1\asec\ of a source in the deep VRO sample (see Tables~1--4 and Figure~2). We would conservatively expect a total of 0.15 spurious X-ray sources at this threshold. The X-ray detected sources in the moderate-depth VRO sample had matching distances of $<0.8$\asec\ although the X-ray detected source in the deep VRO sample had a matching distance of 0.9\asec. This larger matching distance is not unreasonable given the difficulty of source centroid determination when a low number of counts are measured. We inspected these low-significance sources to verify that they are not caused by ``cosmic-ray afterglows'' (\chandra\ X-ray Center 2000, private communication) or background fluctuations.

With these data we have detected X-ray emission from $21^{+12}_{-8}$\% of the \hbox{$HK^{\prime}\le20.4$} VRO population and $11^{+26}_{-9}$\% of the \hbox{$K\le22$} VRO population (see Table~1). All but two of these X-ray detected sources are optically faint with $I\ge24$. Three of the six extreme VROs (i.e.,\ $I-K\ge5$) in the moderate-depth VRO sample are detected with X-ray emission. Two of these sources are extremely faint X-ray sources and have irregular optical morphologies, suggesting they may be dust-enshrouded galaxies (see \S3.4.1).

\subsection{Multi-wavelength counterparts and source redshifts}

In addition to the extremely deep X-ray and moderately deep optical-to-near-IR coverage, the reduced Hawaii Flanking-Field area has the deepest feasible optical spectroscopy (e.g.,\ Cohen \etal 2000; Dawson \etal 2001), deep 1.4 and 8.5~GHz radio observations (Richards \etal 1998; Richards 2000), and shallow sub-millimeter observations (Borys \etal 2001).\footnote{The reduced Hawaii Flanking-Field area also has additional optical coverage (e.g.,\ Hogg \etal 2000); however, we found the Barger \etal (1999) images to provide superior resolution and depth.} A fraction of this region also has deep 15~$\mu$m and sub-millimeter coverage (Aussel \etal 1999a; Hughes \etal 1998; Barger \etal 2000), and the central HDF-N region has some of the deepest multi-wavelength observations taken to date (see Ferguson, Dickinson, \& Williams 2000). We cross-correlated our VROs to the sources in these catalogs using a matching radius of 1\asec\ with the exception of the {\it ISOCAM} catalog of Aussel \etal (1999a) and the {\it SCUBA} catalogs of Hughes \etal (1998) and Borys \etal (2001); see Tables~2--4. Given the larger uncertainty of the {\it ISOCAM} and {\it SCUBA} source positions, we used matching radii of 3\asec\ and 10\asec\ respectively; with these matching distances there is a non-negligible probability of mis-associations, and we would expect 0.5 spurious {\it ISOCAM} matches and 0.6 spurious {\it SCUBA} matches. As all the matches to {\it ISOCAM} sources were those included in the {\it ISOCAM} supplementary catalog and, with the exception of CXOHDFN J123647.9+621019, all had weak emission ($f_{15\rm{\mu m}}<100$~$\mu$Jy), there is also an additional possibility that some of the {\it ISOCAM} sources are spurious (see Aussel \etal 1999b). With the exception of the tentative (2.2$\sigma$) sub-millimeter counterpart of CXOHDFN J123701.6+621145 (see \S3.4.1), we found no matches to {\it SCUBA} sources.

All of the sources in the deep VRO sample have photometric redshifts in the Fernandez-Soto \etal (1999) catalog of $z=$~1.0--1.8 (see Table~3); this redshift range is consistent with that found in other VRO studies to date (e.g.,\ Cimatti \etal 1999; Martini 2001). The comparitively blue near-IR colors of these sources ($J-K<2.3$) are consistent with those expected for elliptical galaxies at $z=$~1--2 (Pozzetti \& Mannucci 2000), and the best fitting spectral energy distribution (SED) template for the majority of these sources is that of an elliptical galaxy.

We are unable to provide reliable photometric redshifts for the sources in the moderate-depth VRO sample and the intermediate-depth X-ray detected VRO sample due to a limited number of optical-to-near-IR bands.

%
%%%%%%%%%%%%%%%%%%%%%%%%%%%%%%%%%%%%%%%%%%%%%%%%%%%%%%%%%%%%%%%%%%%%%%
\section{Constraints on the nature of VROs}
%%%%%%%%%%%%%%%%%%%%%%%%%%%%%%%%%%%%%%%%%%%%%%%%%%%%%%%%%%%%%%%%%%%%%%
%

In this section we provide constraints on the nature of VROs from an analysis of their multi-wavelength properties, focusing on the X-ray emission characteristics of VROs. As most of our X-ray detected VROs do not have redshifts we have also defined an X-ray detected VRO literature sample which includes all X-ray detected VROs from the literature with spectroscopic or photometric redshifts (Fabian \etal 2000; Barger \etal 2001; Cowie \etal 2001; Crawford \etal 2001; Lehmann \etal 2001); this sample is clearly not complete and is only included to provide a comparison to the properties of our VROs.

\subsection{Optical-to-near-IR constraints}

Deep X-ray surveys have shown a correlation between the optical faintness and optical-to-near-IR color of X-ray sources (e.g.,\ Hasinger \etal 1999; Giacconi \etal 2001; Lehmann \etal 2001). In Paper VI we showed that this correlation exists for the X-ray sources in the reduced Hawaii Flanking-Field area with $>$99.99\% confidence. In Figure~3a we plot the mean $I-K$ colors for X-ray sources with $I\le23$ showing this correlation. Clearly the majority of the X-ray detected VROs continue this trend, suggesting they are the extension of the X-ray source population to very red optical-to-near-IR colors. The only X-ray detected VROs that do not follow this trend are the two $I-K\ge5.5$ VROs which may be dust-enshrouded galaxies and different to the majority of the X-ray source population (see \S3.4.1). In Figure~3b we show the mean redshifts for X-ray sources with $I\le23$ and compare them to those expected for $M^{*}_{I}$ E, Sa and Sc galaxies; we determined $M^{*}_{I}$ using the $M^{*}_{B}$ magnitudes given in Folkes \etal (1999) and the $B-I$ colors given in Mannucci \etal (2001). The redshifts of our X-ray undetected VROs are $z\approx$~1--2 (see Table~3), consistent with the few known spectroscopic and photometric redshifts of VROs to date (e.g.,\ Cimatti \etal 1999; Martini 2001; Willott, Rawlings \&
%
%%%%%%%%%%%%%%%%%%%%%%%%%%%%%%%%%%%%%%%%%%%%%%%%%%%%%%%%%%%%%%%%%%%%%%
% 3 Redshifts, magnitudes and colors for the VROs
%%%%%%%%%%%%%%%%%%%%%%%%%%%%%%%%%%%%%%%%%%%%%%%%%%%%%%%%%%%%%%%%%%%%%%
%
\begin{figure*}
\centerline{\includegraphics[angle=-90,width=12.0cm]{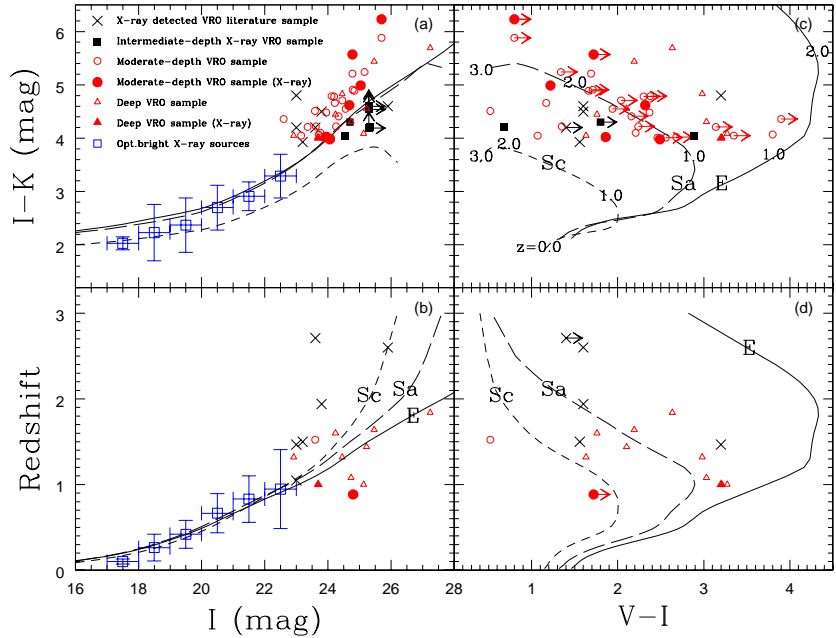}}
\figcaption{Redshifts, $I$-band magnitudes and optical-to-near-IR colors of VROs. The open squares show the average $I-K$ color or spectroscopic redshift for the $I<23$ \chandra\ sources (see Paper~VI); the width of each magnitude bin is shown as bars in the $x$-axis direction. The filled circles, filled triangles and filled squares indicate X-ray detected sources from the moderate-depth VRO sample, the deep VRO sample, and the intermediate-depth X-ray detected VRO sample but not in the moderate-depth VRO sample, respectively. The open circles, open triangles and crosses indicate VROs from the moderate-depth VRO sample, the deep VRO sample, and the X-ray detected VRO literature sample, respectively (see \S2 and \S3.1). The solid, long-dashed and short-dashed curves are the redshift tracks of $M_{I}=-22.5$ E and Sa galaxies and an $M_{I}=-22$ Sc galaxy; these magnitudes correspond to those expected for an $M_{I}^{*}$ galaxy in our assumed cosmology. The galactic $K$-corrections were taken from Poggianti (1997), and the $z\approx$~0 colors were taken from Mannucci \etal (2001). The lower panels only show those VROs with redshifts.}
\label{fig:redshift1}
\end{figure*}

%%%%%%%%%%%%%%%%%%%%%%%%%%%%%%%%%%%%%%%%%%%%%%%%%%%%%%%%%%%%%%%%%%%%%%

\noindent Blundell 2001); on average these sources appear to reside in approximately $M^{*}_{I}$ galaxies. The redshifts of the VROs in the X-ray detected VRO literature sample cover $z=$~1--3, similar to that found for X-ray undetected VROs.

We can also use the $V$-band observations to place further constraints on the properties of X-ray detected VROs. In Figure~3c we show the $V-I$ versus $I-K$ colors of all the VROs and compare them to those expected for different galaxy types.\footnote{We converted the {\it HST} $F606W$-band photometry to the standard $V$ band using $V=0.16+AB_{606}+0.40(AB_{606}-AB_{814})$ (Cowie \etal 1999).} All of the VROs in the deep VRO sample have $V$-band detections, and almost all lie within the region between the Sa and E galaxy tracks.\footnote{We have applied $K$-corrections to these galaxy tracks using the $K$-correction factors given in Poggianti (1997). However, we have not applied evolutionary corrections; hence these tracks should be considered extreme in the sense that a typical $z=1$ elliptical should have bluer colors than those shown here. Since our analysis is not sensitive to the accuracy of these tracks, our overall conclusions will not change even if the true evolutionary corrections are found to be large.} The only X-ray detected source in the deep VRO sample appears to be an elliptical galaxy in the {\it HST} images and lies between the predicted tracks for $z\approx$~1 Sa and E galaxies (see \S3.2.2 and Figures~3c and 3d). Due to the shallower $V$-band coverage of the moderate-depth VRO sample, only 15 of the 29 sources are detected, restricting the constraints we can place on the host galaxy types of these VROs. With the exception of the two $I-K>5.5$ VROs, which are very faint in the optical band and may be dust-enshrouded galaxies (see \S3.4.1), all of the X-ray detected VROs are also detected in the $V$-band; however, the average $V-I$ colors of these sources are consistent with those found for the X-ray undetected VROs with $V$-band detections. Unfortunately the depth and resolution of these images are not sufficient to determine accurately the host-galaxy types of these sources.

\subsection{X-ray constraints for X-ray detected VROs}

Key distinctions between AGN and ``normal'' galaxies can be made at X-ray energies. AGN are generally distinguishable by having moderate-to-high X-ray luminosities, high X-ray-to-optical flux ratios, and, in the case of obscured AGN, flat X-ray spectral slopes.\footnote{The flat X-ray spectral slopes of obscured AGN are due to absorption and are not intrinsic.} In Figure~4a we show the hard-band flux versus the X-ray band ratio, defined as the ratio of hard-band to soft-band counts, for all the X-ray detected VROs. The range of spectral slopes for the nine VROs detected in the hard band is $-0.3<\Gamma<1.5$, and the origin of the X-ray emission from these sources is almost certainly obscured AGN activity (see \S3.2.1). There are, however, four VROs undetected in the hard band; we argue below that at least some of these sources are not obscured AGN (see \S3.2.2).

\subsubsection{Hard-band detected VROs}

The constraints on the X-ray spectral slope for the nine hard-band detected VROs are poor in most cases. However, we can derive an average X-ray spectral slope by stacking the individual sources using the stacking technique presented in \S3.3 of Paper VI. We find a mean band ratio of $0.94\pm0.07$, corresponding to a mean spectral slope of $\Gamma\approx$~0.9. This is very similar to that found for the optically faint X-ray source population (Paper VI), suggesting that these X-ray detected VROs are simply the extension of the optically faint X-ray source population to very red optical-to-near-IR colors. Such a flat X-ray spectral slope is strongly indicative of obscured AGN activity and, assuming the observed X-ray spectral slope is due to obscuration of $\Gamma=2.0$ power-law emission, corresponds to an intrinsic absorption column density at \hbox{$z=$~1--3} (the probable redshift range of these sources; see \S3.1) of \hbox{$N_{\rm H}\approx$~(5--35)$\times10^{22}$ cm$^{-2}$}. Three of the VROs are undetected in the soft band and have inverted spectra ($\Gamma<0.0$); these characteristics can be indicative of Compton-thick absorption. We can derive an average spectral slope for these sources by stacking the individual sources in the same manner as performed above; this gives a mean band ratio of $4.2^{+1.3}_{-1.0}$, corresponding to a mean spectral slope of $\Gamma\approx$~--0.4 and an average intrinsic column-density at \hbox{$z=$~1--3} of \hbox{$N_{\rm H}\approx$~(0.2--1.2)$\times10^{24}$ cm$^{-2}$}. The upper-limit absorption estimate is just below that expected for Compton-thick absorption ($1.5\times10^{24}$ cm$^{-2}$). However, we note that this column density range should be considered a lower limit since additional components other than the intrinsic AGN emission (e.g.,\ scattered AGN emission and star-formation) can raise the soft-band flux.

%
%%%%%%%%%%%%%%%%%%%%%%%%%%%%%%%%%%%%%%%%%%%%%%%%%%%%%%%%%%%%%%%%%%%%%%
% 4 Hardness ratios
%%%%%%%%%%%%%%%%%%%%%%%%%%%%%%%%%%%%%%%%%%%%%%%%%%%%%%%%%%%%%%%%%%%%%%
%

\centerline{\includegraphics[angle=0,width=9.0cm]{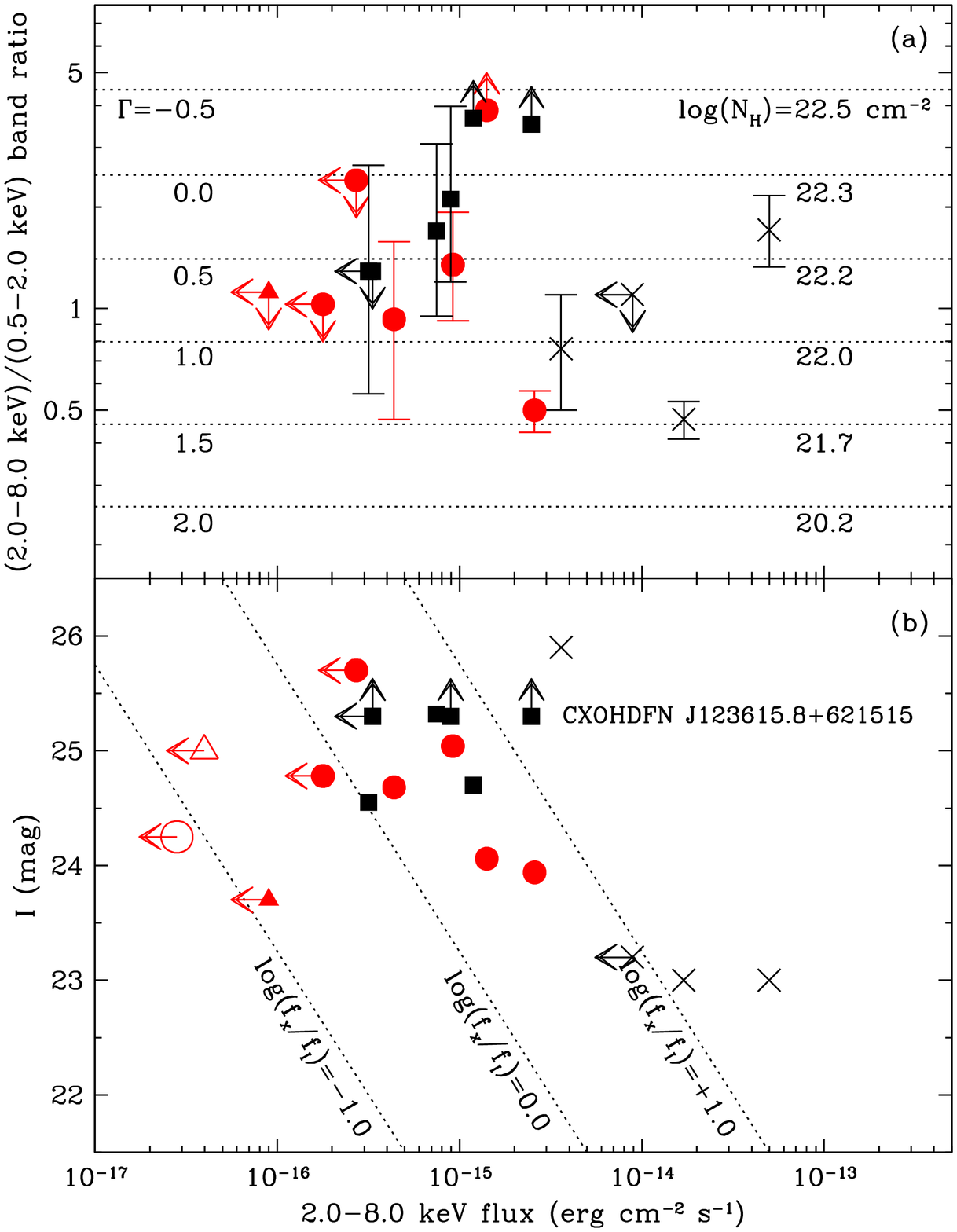}}
\figcaption{Hard-band flux versus (a) X-ray band ratio, defined as the ratio of hard-band to soft-band count rate, and (b) $I$-band magnitude. The symbols have the same meaning as in Figure~3. The open circles and open triangles indicate the stacking-analysis results for the X-ray undetected sources from the moderate-depth VRO sample and deep VRO sample, respectively.  We converted the reported X-ray fluxes and count-rates in the X-ray detected VRO literature sample to the energy bands used here using the AO2 version of {\sc pimms} (Mukai 2000). The equivalent photon indices ($\Gamma$) are shown on the left-hand side of the top panel, and the equivalent column densities ($N_H$) for a $\Gamma$=~2.0 power-law source at $z\approx0$ are shown on the right-hand side of the top panel. These values were determined using {\sc pimms} (Mukai 2000) and {\sc xspec} version 11.0.1 (Arnaud 1996). A possible obscured QSO (CXOHDFN J123615.8+621515) is indicated (see \S3.2.1).}
\label{fig:luminosities}

%%%%%%%%%%%%%%%%%%%%%%%%%%%%%%%%%%%%%%%%%%%%%%%%%%%%%%%%%%%%%%%%%%%%%%

Another key signature of AGN activity is luminous X-ray emission. In determining source luminosities we have corrected for apparent intrinsic absorption in each VRO by assuming that the observed X-ray spectral slope is due to obscuration of $\Gamma=2.0$ power-law emission; the equivalent $z\approx$~0 absorbing column densities for a given band ratio are given in Figure~4a. For those sources that do not have a measured redshift we have assumed $z=1$ and $z=3$. The estimated rest-frame unabsorbed 0.5--8.0~keV luminosities of all the X-ray detected VROs are given in Table~4. The range in X-ray luminosity for the hard-band detected VROs is $\approx$~3$\times10^{42}$ erg~s$^{-1}$ to $\approx$~5$\times10^{44}$ erg~s$^{-1}$, and two sources (CXOHDFN J123615.8+621515 and CXOHDFN J123714.0+620916) would be obscured QSOs (i.e.,\ $>3\times10^{44}$ erg~s$^{-1}$) at $z\approx$~3; see also Paper VI. As star-formation activity is rarely more luminous than $10^{42}$ erg~s$^{-1}$ in the X-ray band (e.g.,\ Moran, Lehnert, \& Helfand 1999 and references therein), it is probable that all the hard-band detected VROs are obscured AGN; these results are consistent with those found in Paper II. If we assume that we have detected all luminous AGN sources in the moderate-depth VRO sample then the fraction of such sources within the $K\le20.1$ VRO population is $14^{+11}_{-7}$\%.

Further evidence that these sources host AGN activity is found from their X-ray-to-optical flux ratios (see Figure~4b). With the exception of one source, all of the VROs detected in the hard band lie within the boundaries expected for typical AGN (i.e.,\ $-1<\log{({{f_{\rm X}}\over{f_{\rm I}}})}<1$; Akiyama \etal 2000; Paper VI). The only source found to lie outside these boundaries is CXOHDFN J123615.8+621515. This source is optically blank ($I>25.3$) and has $\log{({{f_{\rm X}}\over{f_{\rm I}}})}>1$, an inverted X-ray spectral slope ($\Gamma<$~--0.3) and a $z=3$ rest-frame unabsorbed 0.5--8.0~keV luminosity of $\approx$~4$\times10^{44}$ erg~s$^{-1}$. This source may be an obscured QSO; see \S6.2 in Paper VI for further discussion of this source.

\subsubsection{Hard-band undetected VROs}

The four X-ray detected VROs that are undetected in the hard band are also faint in the soft band, limiting the X-ray spectral constraints we can provide. We can, however, derive an average X-ray spectral slope by stacking the individual sources using the technique given in \S3.2.1. This still does not provide an average detection in the hard band, although it places a constraint on the average band ratio of $<0.56$, corresponding to $\Gamma>1.4$. This X-ray spectral slope is considerably steeper than that found for the hard-band detected VROs, suggesting the X-ray emission mechanisms of these sources are different. Since we have redshift determinations for two of these VROs (CXOHDFN J123651.9+621225 and CXOHDFN J123701.6+621145) we can provide better constraints on their X-ray emission mechanism.

CXOHDFN J123651.9+621225 (optical identification 123652.0+621226) lies within the HDF-N itself and has a photometric redshift in the Fern\'andez-Soto \etal (1999) catalog of $z=1.0$; three independent photometric redshift studies also place this galaxy at $z=$~1.0--1.2 (see Hughes \etal 1998). We made our own determination using the publicly available photometric redshift code {\sc hyperz} Version 1.1 (Bolzonella, Miralles, \& Pell\'o 2000).\footnote{The {\sc hyperz} code is available at http://webast.ast.obs-mip.fr/hyperz/.} In performing the photometric redshift fitting, we used only the Bruzual \& Charlot (1993) spectral templates and allowed up to 1.6 mags of visual extinction. The best-fitting model was found to be a $z=1.00^{+0.02}_{-0.06}$ elliptical galaxy with an age of 6.5~Gyr and 0.3 mags of visual extinction; the uncertainty in the redshift determination corresponds to the 90\% confidence level. Indeed, based on optical morphology, this source is clearly an elliptical galaxy and has $V-I$ versus $I-K$ colors consistent with that of an elliptical at $z\approx$~1 (see Figures~3c and 3d). Assuming $\Gamma=2.0$, the rest-frame 0.5--2.0~keV luminosity is $\approx$~9$\times10^{40}$ erg~s$^{-1}$, and the rest-frame 0.5--8.0~keV luminosity is $<3\times10^{41}$ erg~s$^{-1}$. Furthermore, with a $K$-corrected $B$-band magnitude of $M_B=-20.7$, the X-ray emission from this source is consistent with that found for local elliptical galaxies (e.g.,\ O'Sullivan, Forbes, \& Ponman 2001; S.~Pellegrini 2001, private communication).

% MB=-20.7 corresponds to $L_B$=1.6$\times10^{43}$ erg~s$^{-1}$.

CXOHDFN J123701.6+621145 (optical identification 123701.5+621146) is an optically faint $\mu$Jy radio source (Richards \etal 1999) and has a spectroscopic redshift of $z=0.884$, based on a single emission line assumed to be [OII]~$\lambda$3727 (Cohen \etal 2000).\footnote{The SED of CXOHDFN J123701.6+621145 provides corroborating evidence for this redshift (see \S3.4.1).} Adopting $\Gamma=2.0$, the rest-frame 0.5--2.0~keV luminosity is $\approx$~1$\times10^{41}$ erg~s$^{-1}$, and the rest-frame 0.5--8.0~keV luminosity is $\approx$~4$\times10^{41}$ erg~s$^{-1}$. As the irregular morphology of this source (van den Bergh \etal 2000) clearly indicates that it is not an elliptical galaxy, the X-ray emission is probably produced by either luminous star-formation or low-luminosity AGN activity; we provide further constraints in \S3.4.1.

We do not have detailed constraints for the other two hard-band undetected VROs, CXOHDFN J123639.6+620936 and CXOHDFN J123642.8+620958. However, their irregular $HK^{\prime}$-band morphologies and extremely faint optical magnitudes ($I\simgt25$) suggest they may be similar to CXOHDFN J123701.6+621145 and not normal elliptical galaxies. The extreme optical-to-near-IR color of CXOHDFN J123642.8+620958 ($I-K=6.2$) and the lower limit optical-to-near-IR color of CXOHDFN J123639.6+620936 ($I-K>4.5$) provide further support for this interpretation. 

If we assume the hard-band undetected sources in the moderate-depth VRO sample with irregular morphologies are luminous star-forming galaxies, the fraction of such sources within the $K\le20.1$ VRO population is $7^{+8}_{-5}$\%. Clearly, this fraction should be used with caution since we have not proved that these sources are luminous star-forming galaxies, and a large number of such sources may lie below the detection limit of our X-ray observations.

\subsection{X-ray constraints for X-ray undetected VROs}

The majority of our VROs do not have detectable X-ray emission. We can, however, search for evidence of X-ray emission from these sources on average by stacking the individually undetected sources in the same manner as performed in previous studies (e.g.,\ Paper II; Paper IV; Paper VI). We conservatively excluded from the stacking analyses five of the 23 X-ray undetected VROs in the moderate-depth VRO sample and one of the eight X-ray undetected VROs in the deep VRO sample that lay within 10\asec\ of a high-significance X-ray source (i.e.,\ detected with a {\sc wavdetect} false-positive probability threshold of $10^{-7}$) or within 3\asec\ of a low-significance X-ray source (i.e.,\ detected with a {\sc wavdetect} threshold of $10^{-4}$ but undetected with a {\sc wavdetect} threshold of $10^{-7}$); these could be affected by the low-level wings of the \chandra\ point-spread function (see Tables~2 and 3). In all the stacking analyses we used the 30-pixel aperture and restricted ACIS grade set defined in Paper IV. To determine the significance of the stacked X-ray emission we performed Monte-Carlo stacking simulations using the technique given in Appendix A of Paper IV. The Monte-Carlo simulations consisted of 10,000 trials where we stacked 18 random positions (seven random positions in the case of the deep VRO sample) within each VRO sample area using the same photometric aperture as for the stacking analyses. We verified the significances determined from these simulations using Poisson statistics; all significant detections (i.e.,\ $\ge99$\% confidence level) were found to be so with both methods. The results of the stacking analyses are given in Table~5; we have assumed $\Gamma=2.0$ when determining all fluxes and luminosities.

\subsubsection{Results from the moderate-depth VRO sample}

The effective \chandra\ exposure time achieved from the stacking analyses of the moderate-depth VRO sample is 15.8~Ms. We find significant detections ($\ge$99\% confidence level) in the soft and full bands corresponding to average fluxes of $6.0\times10^{-18}$~erg~cm$^{-2}$~s$^{-1}$ and $1.5\times10^{-17}$~erg~cm$^{-2}$~s$^{-1}$, respectively.

We may expect the optically brightest sources to contribute more to the stacked X-ray emission as for a given X-ray-to-optical flux ratio these sources would have brighter X-ray fluxes. To investigate this possibility we performed separate stacking analyses for the $I\le24.1$ and $I>24.1$ VROs; the $I=24.1$ threshold was chosen to give equal numbers of VROs in both sub-samples. The effective \chandra\ exposure times from these stacking analyses are similar (8.1~Ms for the $I\le24.1$ sub-sample and 7.7~Ms for the $I>24.1$ sub-sample) and allow for a relatively unbiased comparison. We find significant detections in the soft and full bands and a tentative (96.4\% confidence level) detection in the hard band for the $I\le24.1$ VROs; there are no significant detections for the $I>24.1$ VROs in any X-ray band (see Table~5). This clearly shows that the $I\le24.1$ VROs are contributing the majority of the stacked X-ray emission, and deeper \chandra\ observations should preferentially detect the optically brightest VROs. Given the average soft-band flux ($8.7\times10^{-18}$~erg~cm$^{-2}$~s$^{-1}$) and mean $I$-band magnitude ($I=23.7\pm0.3$) of the $I\le24.1$ VROs, the mean X-ray-to-optical flux ratio of $\log{({{f_{\rm X}}\over{f_{\rm I}}})}=-1.9$ is consistent with that expected from normal elliptical galaxies. Furthermore, given the mean $I$-band magnitude for the $I>24.1$ sources ($I=24.8\pm0.6$) and assuming the same X-ray-to-optical flux ratio, the expected soft-band flux of $3.3\times10^{-18}$~erg~cm$^{-2}$~s$^{-1}$ is consistent with our non-detection of the $I>24.1$ VROs.

The fainter $I$-band magnitudes of the $I>24.1$ VROs could be due to these sources lying at higher redshifts than the $I\le24.1$ VROs. Although we do not have redshifts for the majority of these VROs, assuming they are $M^{*}_{I}$ elliptical galaxies the average redshifts of the $I\le24.1$ and $I>24.1$ sources are $z=1.1$ and $z=1.4$, respectively. At $z=1.1$, the average rest-frame 0.5--2.0~keV and 0.5--8.0~keV luminosities of the $I\le24.1$ VROs are $5.0\times10^{40}$ erg~s$^{-1}$ and $1.2\times10^{41}$ erg~s$^{-1}$, respectively. This X-ray emission is at the level expected from normal elliptical galaxies, although it could also be produced by star formation and low-luminosity AGN activity. At $z=1.4$, the average rest-frame 0.5--2.0~keV and 0.5--8.0~keV luminosity upper limits of the $I>24.1$ VROs are $<6.2\times10^{40}$ erg~s$^{-1}$ and $<2.1\times10^{41}$ erg~s$^{-1}$, respectively. Both of these X-ray luminosity upper limits are higher than those determined for the $I\le24.1$ VROs and therefore the non-detection of X-ray emission from the $I>24.1$ VROs is consistent with these sources lying at higher redshifts. Deep imaging in multiple optical-to-near-IR bands is required to determine source morphologies and photometric redshifts to test these hypotheses (see \S4.2).

\subsubsection{Results from the deep VRO sample}

The effective \chandra\ exposure time achieved from the stacking analyses of the deep VRO sample is 6.5~Ms. We find a significant detection ($\ge$99\% confidence level) in the soft band corresponding to an average flux of $6.5\times10^{-18}$~erg~cm$^{-2}$~s$^{-1}$. At the mean photometric redshift of these sources ($z=1.5\pm0.2$; see Table~3), the rest-frame 0.5--2.0~keV luminosity is $9\times10^{40}$ erg~s$^{-1}$, the same as that found for the one X-ray detected VRO in the deep VRO sample (CXOHDFN J123651.9+621225 at $z=1.0$; see \S3.2.2). As CXOHDFN J123651.9+621225 is the lowest redshift source in the deep VRO sample, this suggests that the individual non-detections of the other VROs are mostly due to these sources lying at higher redshifts. This result is entirely consistent with that suggested in \S3.3.1 based on the stacking analyses of X-ray undetected sources in the moderate-depth VRO sample, suggesting that deeper \chandra\ observations should detect more examples of X-ray emitting normal elliptical galaxies in the VRO population.

As only 7 VROs were used in these stacking analyses we are unable to investigate distinctions in average X-ray properties with optical brightness or redshift.

\subsection{Multi-wavelength constraints}

\subsubsection{Broad-band SED of CXOHDFN J123701.6+621145}

Most of our VROs are only detected at optical-to-near-IR wavelengths; however, CXOHDFN J123701.6+621145 is remarkable in that it is detected at most of the wavelengths investigated here. This source is an extreme VRO (i.e.,\ $I-K\ge5$), has an irregular morphology (van den Bergh \etal 2000), and has a spectroscopic redshift of $z=0.884$ (Cohen \etal 2000); see also \S3.2.2. The extreme VRO colors and irregular morphology suggest this source may be a member of the dust-enshrouded VRO population.

In Figure~5 we compare the SED of CXOHDFN J123701.6+621145 to that of the archetypal nearby dusty starburst galaxy Arp~220. The fluxes of Arp~220 have been adjusted to $z=0.884$ to show the relative luminosities of both sources from radio to X-ray wavelengths; however, to produce the fit shown, we had to reduce the luminosity of the Arp~220 model by a factor of 2.2. Clearly CXOHDFN J123701.6+621145 has a similar SED to Arp~220. The bolometric luminosity of CXOHDFN J123701.6+621145 is lower than that of Arp~220, although the X-ray luminosity is $\approx2.5$ times higher (see \S3.2.2). The X-ray emission from Arp~220 is consistent with that produced by both a star-formation component and a hard X-ray power-law component (Iwasawa \etal 2001); this latter component could be due to AGN emission or a population of X-ray binaries. We do not have sufficient X-ray information to determine the origin of the X-ray emission from CXOHDFN J123701.6+621145; although we note that it is less luminous than that found for NGC~3256 (Moran \etal 1999; Lira \etal 2001), the most X-ray luminous local starburst.

%
%%%%%%%%%%%%%%%%%%%%%%%%%%%%%%%%%%%%%%%%%%%%%%%%%%%%%%%%%%%%%%%%%%%%%%
% 5 Comparison of SEDs
%%%%%%%%%%%%%%%%%%%%%%%%%%%%%%%%%%%%%%%%%%%%%%%%%%%%%%%%%%%%%%%%%%%%%%
%

\vspace{0.2in}
\centerline{\includegraphics[angle=-90,width=9.0cm]{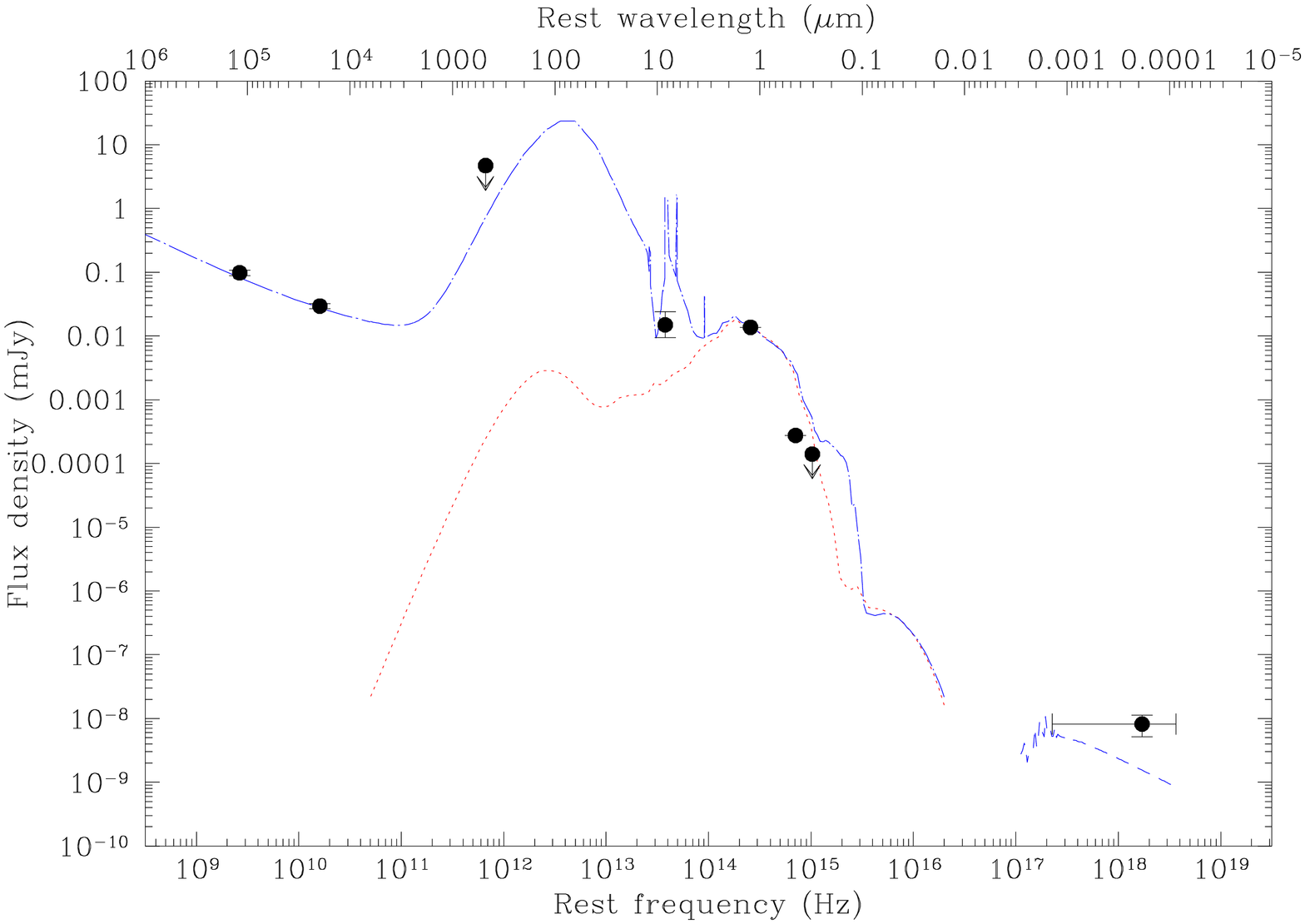}}
\figcaption{A comparison between the multi-wavelength properties of CXOHDFN J123701.6+621145 and the SED of the dusty starburst galaxy Arp 220. The filled circles show the data for CXOHDFN J123701.6+621145. The X-ray and optical-to-near-IR data are taken from this paper while the radio, sub-millimeter and infrared data are taken from Richards (2000), Barger, Cowie, \& Richards (2000), and Aussel \etal (1999a), respectively. Observational data are not shown for Arp 220; instead the best-fit model from Silva \etal (1998) is shown for the radio-to-optical wavelengths (dot-dashed curve), and the best-fit model from Iwasawa \etal (2001) is shown for X-ray energies (dashed curve). These models have been corrected to that expected from Arp 220 at $z=0.884$, although to achieve the fit shown we decreased the luminosity of Arp 220 by a factor of 2.2. The model of a 5~Gyr old passively evolving giant elliptical galaxy taken from Silva \etal (1998) is plotted (dotted curve) to demonstrate the expected 15~$\mu$m emission from starlight.}
\label{fig:redshift1}
\vspace{0.2in}

%%%%%%%%%%%%%%%%%%%%%%%%%%%%%%%%%%%%%%%%%%%%%%%%%%%%%%%%%%%%%%%%%%%%%%

The similarity of the $I-K$ colors and X-ray properties of CXOHDFN J123639.6+620936 and CXOHDFN J123642.8+620958 (see \S3.2.2) with those of CXOHDFN J123701.6+621145 suggests they may also be dust-enshrouded galaxies.

\subsubsection{Radio detections of VROs}

Five VROs have radio counterparts, four of which are also detected at X-ray energies. This high matching fraction between radio and X-ray sources is consistent with that found in other studies (i.e.,\ Paper II; Paper IV; Bauer \etal 2001a) suggesting an association between the production of X-ray and radio emission. Only three of the five VROs with radio counterparts have well-defined radio spectral slopes; two sources (123641.4+621143 and CXOHDFN J123707.2+621408) have flat radio spectra (i.e.,\ $\alpha_R\approx~0.3$) while a third source (CXOHDFN J123701.6+621145) has a steep radio spectrum ($\alpha_R=0.82^{+0.09}_{-0.10}$) and is described in \S3.4.1. The origin of the radio emission in the sources with flat radio spectra can be either AGN or recent starburst activity (e.g.,\ Condon 1992; Richards 2000). The flat X-ray spectrum of CXOHDFN J123707.2+621408 favors the former case for this object. The source 123641.4+621143 is undetected in all the X-ray bands; given a full-band upper limit of $\approx2\times10^{-16}$~erg~cm$^{-2}$~s$^{-1}$, its rest-frame 0.5--8.0~keV luminosity can be as high as $3\times10^{42}$ erg~s$^{-1}$ at $z=1.524$ (Dawson \etal 2001). As this source has a disk-like morphology, it is most likely a star-forming galaxy or a low-luminosity AGN.

\subsubsection{{\it ISOCAM} detections of VROs}

Four of the 14 VROs in the {\it ISOCAM} region of Aussel \etal (1999a) have 15~$\mu$m counterparts, two of which are also detected at X-ray energies (see Tables 2 and 4). This infrared radiation can be due to either dust or starlight emission. In Figure~5, we show the SEDs of both a dusty starburst (Arp~220) and a starlight-dominated galaxy (a giant elliptical galaxy) normalized to the $HK^{\prime}$-band emission of CXOHDFN J123701.6+621145. The predicted 15~$\mu$m emission from starlight is an order of magnitude below that expected from the dusty starburst, and clearly the origin of the infrared radiation in CXOHDFN J123701.6+621145 is dust emission. As CXOHDFN J123701.6+621145 is the brightest {\it ISOCAM}-detected VRO in the $HK^{\prime}$-band, this suggests that starlight emission at 15~$\mu$m for $HK^{\prime}>19.5$ VROs at $z\approx$~1 is $<1\mu$Jy. This is 1--2 orders of magnitude fainter than the 15~$\mu$m flux limit, and the origin of the infrared emission in all the {\it ISOCAM}-detected VROs is likely to be dust emission.

\subsection{VRO clustering}

Wide-area VRO studies have shown that VROs are strongly clustered (e.g.,\ Daddi \etal 2000, 2001; McCarthy \etal 2001). The solid angles covered by our VRO samples are too small to perform a detailed analysis of source clustering; the results of a two-point correlation analysis on the moderate-depth VRO sample with the estimator of Landy \& Szalay (1993) did not show significant evidence for source clustering. 

We also took an alternative approach and searched for VRO overdensities in the vicinities of three known moderate-redshift (i.e.,\ $z\approx$~1) clusters and cluster candidates (see Figure~2). Three VROs lie within the $\approx$~45\asec\ radius region of the cluster-candidate X-ray source CXOHDFN J123620.0+621554 (Bauer \etal 2001b, hereafter Paper IX) when we would expect 0.9 VROs in the case of a uniform distribution. This unusual X-ray source also contains an overdensity of optically faint X-ray sources and optically faint $\mu$Jy radio sources, and it is likely to be a poor cluster at $z\simgt0.7$; see Paper IX for further discussion. We found one VRO within 45\asec\ of the center of the $z=0.85$ cluster ClG 1236+6215 discovered by Dawson \etal (2001) and two VROs (excluding the VROs in the deep VRO sample) within 45\asec\ of the center of the $z=1.01$ cluster/group associated with the FR~I radio galaxy VLA J123644.3+621133 (e.g.,\ Richards \etal 1998; Paper IX). The evidence for VRO clustering within the vicinities of all these sources is not significant although we are limited by statistics. If we combine these results we find six VROs when we would expect 2.7 in the case of a uniform distribution; the evidence for VRO clustering is significant at the 94\% confidence level.

% CXOHDFN J123620.0+621554 probability is 94\%; ClG 1236+6215 is 40\% and VLA J123644.3+621133 is 77\%
% Combining all results together we get 5.85740748667009e-03 which is significance at the 94% confidence level

%
%%%%%%%%%%%%%%%%%%%%%%%%%%%%%%%%%%%%%%%%%%%%%%%%%%%%%%%%%%%%%%%%%%%%%%
\section{Summary and Further work}
%%%%%%%%%%%%%%%%%%%%%%%%%%%%%%%%%%%%%%%%%%%%%%%%%%%%%%%%%%%%%%%%%%%%%%
%

\subsection{Summary of main results}

We have used the 1~Ms CDF-N observation of an $8.4\amin\times8.4\amin$ region within the Hawaii Flanking-Field area to provide constraints on the X-ray emission from Very Red Objects (VROs; $I-K\ge4$). We defined three VRO samples: a moderate-depth VRO sample ($HK^{\prime}\le20.4$) and an intermediate-depth X-ray detected VRO sample ($HK^{\prime}\le21.4$) both covering the 70.3~arcmin$^{2}$ reduced Hawaii Flanking-Field area, and a deep VRO sample ($K\le22$) covering the 5.3~arcmin$^{2}$ region of the HDF-N itself. Our main results are the following:

(i)	We detected X-ray emission from 13 VROs down to 0.5--2.0~keV and 2.0--8.0~keV point-source fluxes of $\approx~3\times10^{-17}$ erg~cm$^{-2}$~s$^{-1}$ and $\approx~2\times10^{-16}$ erg~cm$^{-2}$~s$^{-1}$, respectively. Six ($21^{+12}_{-8}$\%) of the 29 VROs within the moderate-depth VRO sample and one ($11^{+26}_{-9}$\%) of the nine VROs within the deep VRO sample were detected with X-ray emission. See \S2.3 and \S2.4.

(ii)	The $I-K$ color versus $I$-band magnitude of all the $I-K<5.5$ X-ray detected VROs follows the trend found for optically brighter ($I\le23$) X-ray sources, suggesting the VROs are the $z\ge1$ very red analogs of the optically bright X-ray source population. The two $I-K>5.5$ X-ray detected VROs have redder $I-K$ colors than those expected given their $I$-band magnitudes and may be dust-enshrouded VROs. See \S3.1 and \S3.4.1.

(iii)	The hard-band detected VROs have flat X-ray spectral slopes ($\Gamma\approx0.9$) and X-ray properties consistent with those expected from obscured AGN; three of these VROs are undetected in the soft band, and the absorption to the X-ray nucleus in some of these sources may be Compton-thick. Assuming all the hard-band detected VROs lie at $z=$~1--3, they have rest-frame 0.5--8.0~keV unabsorbed luminosities of $\approx$~$3\times10^{42}$ to $5\times10^{44}$ erg~s$^{-1}$; the two brightest sources would be obscured QSOs (i.e.,\ $>3\times10^{44}$~erg~s$^{-1}$) if they lie at $z\approx$~3. Assuming these hard-band detected VROs comprise all the luminous AGNs within the moderate-depth VRO sample, the fraction of such sources in the $HK^{\prime}\le20.4$ VRO population is $14^{+11}_{-7}$\%. Unfortunately, the depth and resolution of the optical-to-near-IR images are not sufficient to determine the host-galaxy types of these sources. See \S3.2.1.

(iv)	The hard-band undetected VROs have steeper X-ray spectral slopes and are more likely to contain less energetic X-ray emission processes such as star formation, low-luminosity AGN, and normal elliptical galaxy emission (i.e.,\ X-ray binary activity and thermal gas emission). One of these sources appears to be a normal elliptical galaxy at $z\approx$~1, while another source is a $z=0.884$ $I-K>5.5$ VRO that has a SED very similar to that of the archetypal dusty starburst galaxy Arp~220. The X-ray emission from this source is likely to be due to either luminous star-formation or low-luminosity AGN activity. The irregular morphologies and optical-to-near-IR colors of the other hard-band undetected VROs are similar to those of the $z=0.884$ extreme VRO, and the X-ray emission from these sources is also likely to be produced by the same processes. Assuming the hard-band undetected sources with irregular morphologies in the moderate-depth VRO sample are luminous star-forming galaxies, the fraction of such sources within the $K\le20.1$ VRO population is $7^{+8}_{-5}$\% or larger. See \S3.2.2 and \S3.4.1.

(v)	Stacking analyses of VROs not individually detected with X-ray emission produced significant detections ($\ge99$\% confidence level) in the 0.5--2.0~keV band for both the moderate-depth and deep VRO samples, and a significant detection in the 0.5--8.0~keV band for the moderate-depth VRO sample. These average X-ray detections are consistent with the emission expected from normal elliptical galaxies, star formation or low-luminosity AGN activity at $z\approx$~1. We found that the optically brightest VROs contribute the majority of this X-ray emission. The simplest explanation of these results is that we have detected the X-ray emission from non-active VROs with low X-ray-to-optical flux ratios [$\log{({{f_{\rm X}}\over{f_{\rm I}}})}\approx$~--2]. This is consistent with expectations if the majority of these VROs are $\approx{M^{*}_{I}}$ elliptical galaxies. See \S3.3.

(vi)	Five VROs have radio counterparts, and four of these are detected with X-ray emission. This high matching rate between radio and X-ray sources is consistent with that found from other deep X-ray and radio studies. See \S3.4.2.

(vii)	Four of the 14 VROs within the {\it ISOCAM} HDF-N region have 15~$\mu$m counterparts, and two of these are detected with X-ray emission. The origin of the infrared radiation from all these {\it ISOCAM} sources is likely to be dust emission. See \S3.4.3.

\subsection{Further work}

Given our results we can provide suggestions for further work to extend the understanding of X-ray emission from VROs. Arguably the most important issue to address is identifying the differences between X-ray detected and X-ray undetected VROs. There are three main approaches that may provide useful constraints in this regard: (i) galactic host types and environments, (ii) redshift distributions, and (iii) large-scale clustering properties.

The determination of galactic host types and environments will be most effectively pursued with high-resolution deep (i.e.,\ $I\approx$~27) optical/near-IR imaging; the excellent angular resolution of {\it HST} is preferable in this regard. As the photometric redshift technique will probably be the only viable method of determining the redshifts for most VROs, deep imaging in multiple optical-to-near-IR bands will be necessary to determine redshift distributions. However, given that many of the X-ray detected VROs are probably AGN, deep optical spectroscopy may successfully detect high equivalent-width emission lines in some sources. It is becoming clear that VROs are strongly clustered although the clustering properties of X-ray detected VROs are almost completely unknown. The advantage of an investigation into the clustering properties of X-ray detected VROs is that it can be undertaken with comparatively moderate depth optical-to-near-IR imaging (i.e.,\ $I\approx$~25). However, to place useful constraints and construct statistically significant samples, imaging observations over a large area (i.e.,\ a few hundred arcmin$^{2}$) would be required.

The 1~Ms \chandra\ observation used in the present study has shown that VROs without obvious AGN activity are probably only detected at faint X-ray fluxes (i.e.,\ 0.5--2.0~keV fluxes $<10^{-16}$~erg~cm$^{-2}$~s$^{-1}$) and are therefore only detectable in extremely deep X-ray surveys. Our ultimate aim with the \hbox{CDF-N} survey is to reach a \chandra\ exposure time of $\approx~5$~Ms. Such an observation should reach 0.5--2.0~keV fluxes of $\approx~6\times10^{-18}$ erg~cm$^{-2}$~s$^{-1}$ and will be deep enough to detect a large number of VROs without obvious AGN activity.

%
%%%%%%%%%%%%%%%%%%%%%%%%%%%%%%%%%%%%%%%%%%%%%%%%%%%%%%%%%%%%%%%%%%%%%%
\section*{Acknowledgments}
%%%%%%%%%%%%%%%%%%%%%%%%%%%%%%%%%%%%%%%%%%%%%%%%%%%%%%%%%%%%%%%%%%%%%%
%

This work would not have been possible without the support of the entire \chandra\ and ACIS teams; we particularly thank P.~Broos and L.~Townsley for data analysis software and CTI correction support.
We thank L.~Pozzetti for interesting discussions, L.~Silva and G.L.~Granato for providing model SEDs, and S.~Dawson, M.~Dickinson, S.~Pellegrini, and P.~Severgnini for providing useful information.
We are grateful to A.~Barger and L.~Cowie for making their optical and near-IR images publicly available and to M.~Bolzonella, J.-M.~Miralles and R.~Pell\'o for making {\sc hyperz} available.
We thank the referee for concise comments.
We acknowledge the financial support of
NASA grant NAS~8-38252 (GPG, PI),
NSF CAREER award AST-9983783 (DMA, CV, FEB, WNB),  
NASA GSRP grant NGT5-50247 and 
the Pennsylvania Space Grant Consortium (AEH), and
NSF grant AST-9900703~(DPS).
C.V. also acknowledges partial support from the Italian Space Agency, under the contract ASI 00/IR/103/AS, and the Ministry for University and Research (MURST) under grant Cofin-00-02-36.

%
%%%%%%%%%%%%%%%%%%%%%%%%%%%%%%%%%%%%%%%%%%%%%%%%%%%%%%%%%%%%%%%%%%%%%%

%
%%%%%%%%%%%%%%%%%%%%%%%%%%%%%%%%%%%%%%%%%%%%%%%%%%%%%%%%%%%%%%%%%%%%%%%%%%%%%%%%%%
% TABLES
%%%%%%%%%%%%%%%%%%%%%%%%%%%%%%%%%%%%%%%%%%%%%%%%%%%%%%%%%%%%%%%%%%%%%%%%%%%%%%%%%%
%

\clearpage

%
%%%%%%%%%%%%%%%%%%%%%%%%%%%%%%%%%%%%%%%%%%%%%%%%%%%%%%%%%%%%%%%%%%%%%%
% TABLE 1: VRO statistics
%%%%%%%%%%%%%%%%%%%%%%%%%%%%%%%%%%%%%%%%%%%%%%%%%%%%%%%%%%%%%%%%%%%%%%
%

%
%%%%%%%%%%%%%%%%%%%%%%%%%%%%%%%%%%%%%%%%%%%%%%%%%%%%%%%%%%%%%%%%%%%%%%
% TABLE 1
%%%%%%%%%%%%%%%%%%%%%%%%%%%%%%%%%%%%%%%%%%%%%%%%%%%%%%%%%%%%%%%%%%%%%%
%

\begin{deluxetable}{lcccccccc}
%\rotate
\tablecolumns{9}
\tabletypesize{\scriptsize}
\tablewidth{0pt}
\tablecaption{VRO sample definitions and basic properties}
\tablehead{
\colhead{VRO} &
\colhead{Magnitude} &
\colhead{} &
\colhead{VRO} & 
\colhead{} &
\colhead{Source} &
\colhead{} &
\colhead{X-ray} &
\colhead{Source} \\
\colhead{Sample} &
\colhead{Limit$^{\rm a}$} &
\colhead{Area$^{\rm b}$} &
\colhead{Definition} &
\colhead{$N^{\rm c}$} &
\colhead{Density$^{\rm d}$} &
\colhead{$N_X$$^{\rm e}$} &
\colhead{Fraction$^{\rm f}$} &
\colhead{Density$^{\rm g}$}}
\startdata
      Moderate-depth&     $HK^{\prime}\le20.4$&70.3&                $I-K\ge4$& 29&          $1500^{+350}_{-250}$&       $  6^{\rm h}$&             $21^{+12}_{-8}$\%&           $300^{+200}_{-100}$\\
\bigskip
                    &                         &                         &                $I-K\ge5$&  6&           $300^{+200}_{-100}$&       $  3^{\rm h}$&            $50^{+49}_{-27}$\%&           $150^{+150}_{-100}$\\
\bigskip
                                    Deep&               $K\le22.0$& 5.3&                $I-K\ge4$&  9&        $6100^{+2800}_{-2000}$&       $  1^{\rm h}$&             $11^{+26}_{-9}$\%&          $700^{+1550}_{-550}$\\
       Intermediate-depth X-ray detected&     $HK^{\prime}\le21.4$&70.3&                $I-K\ge4$& 10&           $500^{+200}_{-150}$&       $ 10^{\rm i}$&                         100\%&           $500^{+200}_{-150}$\\
\enddata

\tablenotetext{a}{For the $HK^{\prime}$-selected samples, the conversion to $K$-band magnitudes is $K$=$HK^{\prime}-(0.13+0.05(I-K))$ (Barger et~al. 1999).}
\tablenotetext{b}{Sample area (arcmin$^{2}$).}
\tablenotetext{c}{Number of VROs.}
\tablenotetext{d}{Source density of VROs (deg$^{-2}$).}
\tablenotetext{e}{Number of X-ray detected VROs.}
\tablenotetext{f}{Fraction of X-ray detected VROs.}
\tablenotetext{g}{Source density of X-ray detected VROs (deg$^{-2}$).}
\tablenotetext{h}{X-ray sources detected by {\sc wavdetect} with false-detection probability threshold $10^{-4}$.}
\tablenotetext{i}{X-ray sources detected by {\sc wavdetect} with false-detection probability threshold $10^{-7}$.}

\end{deluxetable}

%
%%%%%%%%%%%%%%%%%%%%%%%%%%%%%%%%%%%%%%%%%%%%%%%%%%%%%%%%%%%%%%%%%%%%%%
% TABLE 2: Moderate-depth VROs
%%%%%%%%%%%%%%%%%%%%%%%%%%%%%%%%%%%%%%%%%%%%%%%%%%%%%%%%%%%%%%%%%%%%%%
%

%
%%%%%%%%%%%%%%%%%%%%%%%%%%%%%%%%%%%%%%%%%%%%%%%%%%%%%%%%%%%%%%%%%%%%%%
% TABLE 2
%%%%%%%%%%%%%%%%%%%%%%%%%%%%%%%%%%%%%%%%%%%%%%%%%%%%%%%%%%%%%%%%%%%%%%
%

\begin{deluxetable}{llccccc}
%\rotate
\tablecolumns{7}
\tabletypesize{\scriptsize}
\tablewidth{0pt}
\tablecaption{Moderate-depth VRO sample}
\tablehead{
\multicolumn{2}{c}{Optical coordinates} &
\multicolumn{4}{c}{Optical magnitudes} &
\colhead{} \\
\multicolumn{1}{c}{$\alpha_{2000}$}&
\multicolumn{1}{c}{$\delta_{2000}$}&{$I^{\rm a}$}&{$HK^{\prime \rm a}$}&{$I-K^{\rm b}$}&{$V^{\rm a}$}&{Other $\lambda^{\rm c}$}}
\startdata
         12 36 13.36&                                            +62 17 15.0$^{\rm d}$&                24.8&                19.9&                 5.2&                26.5&                    \\
         12 36 14.55&                                            +62 13 44.0$^{\rm d}$&                23.8&                20.0&                 4.1&                26.0&                    \\
         12 36 18.73&                                            +62 16 01.6$^{\rm d}$&                24.5&                20.0&                 4.7&             $>$26.5&                    \\
         12 36 20.49&                                                      +62 12 37.8&                24.7&                20.4&                 4.6&                27.0&                   X\\
         12 36 20.93&                                                      +62 16 07.3&                23.9&                20.2&                 4.0&             $>$26.5&                    \\
\\
         12 36 24.14&                                            +62 09 11.5$^{\rm d}$&                24.6&                20.3&                 4.6&             $>$26.5&                    \\
         12 36 25.32&                                            +62 10 35.8$^{\rm d}$&                24.0&                20.2&                 4.1&                25.0&                    \\
         12 36 27.63&                                            +62 15 33.9$^{\rm d}$&                24.0&                19.8&                 4.5&                26.3&                    \\
         12 36 30.75&                                                      +62 15 13.5&                24.3&                20.2&                 4.4&             $>$26.5&                    \\
         12 36 34.24&                                            +62 14 00.4$^{\rm d}$&                24.1&                19.6&                 4.8&             $>$26.5&                    \\
\\
         12 36 35.73&                                            +62 10 32.9$^{\rm d}$&                23.7&                19.9&                 4.1&                27.5&                    \\
         12 36 38.83&                                            +62 14 51.7$^{\rm d}$&                25.2&                20.2&                 5.2&             $>$26.5&                   I\\
         12 36 40.89&                                            +62 15 04.9$^{\rm d}$&                23.8&                20.1&                 4.0&    $\cdots^{\rm e}$&                    \\
         12 36 41.37&                                                      +62 11 29.7&                22.6&                18.5&                 4.4&    $\cdots^{\rm e}$&                    \\
         12 36 41.41&                                            +62 11 42.6$^{\rm d}$&                23.6&                19.4&                 4.5&                24.1&                 R,Z\\
\\
         12 36 42.89&                                                      +62 09 58.9&                25.7&                19.8&                 6.2&             $>$26.5&                   X\\
         12 36 43.91&                                            +62 16 21.0$^{\rm d}$&                24.2&                19.9&                 4.7&                25.4&                    \\
         12 36 44.12&                                            +62 12 44.9$^{\rm d}$&                24.2&                19.7&                 4.8&             $>$26.5&                    \\
         12 36 53.79&                                            +62 11 18.0$^{\rm d}$&                24.8&                20.2&                 4.9&             $>$26.5&                    \\
         12 36 58.80&                                                      +62 10 22.6&                24.1&                20.4&                 4.0&                26.5&                   X\\
\\
         12 37 01.57&                                                      +62 11 46.1&                24.8&                19.5&                 5.6&             $>$26.5&             X,R,I,Z\\
         12 37 03.75&                                                      +62 13 52.0&                23.4&                19.5&                 4.2&    $\cdots^{\rm e}$&                   I\\
         12 37 05.67&                                            +62 11 13.7$^{\rm d}$&                25.7&                20.1&                 5.9&             $>$26.5&                    \\
         12 37 07.20&                                                      +62 14 08.1&                25.0&                20.4&                 5.0&                26.3&                 X,R\\
         12 37 11.17&                                            +62 09 33.1$^{\rm d}$&                23.1&                19.4&                 4.1&             $>$26.5&                    \\
\\
         12 37 12.08&                                            +62 10 50.8$^{\rm d}$&                23.6&                19.7&                 4.2&                25.0&                    \\
         12 37 12.46&                                            +62 09 44.4$^{\rm d}$&                24.8&                20.2&                 4.9&             $>$26.5&                    \\
         12 37 14.04&                                                      +62 09 16.8&                23.9&                20.2&                 4.0&                25.8&                   X\\
         12 37 23.88&                                                      +62 14 00.5&                24.3&                20.3&                 4.2&             $>$26.5&                    \\
\enddata

\tablenotetext{a}{Vega-based magnitude.}
\tablenotetext{b}{Calculated $I-K$ color. The $HK^{\prime}$-band magnitudes are converted to $K$-band magnitudes following Barger et~al. (1999); see Table~1.}
\tablenotetext{c}{Source has multi-wavelength counterpart. (X) X-ray counterpart, see Table~4; (R) Radio counterpart (Richards et~al. 1998; Richards 2000); (I) Mid-infrared counterpart (Aussel et~al. 1999a); (Z) Spectroscopic redshift (Cohen et~al. 2000; Dawson et~al. 2001).}
\tablenotetext{d}{Source used in stacking analysis; see \S3.3.1.}
\tablenotetext{e}{Source is detected in the $V$-band, but a magnitude could not be accurately determined due to the close proximity of another source.}

\end{deluxetable}

%
%%%%%%%%%%%%%%%%%%%%%%%%%%%%%%%%%%%%%%%%%%%%%%%%%%%%%%%%%%%%%%%%%%%%%%
% TABLE 3: Deep VROs
%%%%%%%%%%%%%%%%%%%%%%%%%%%%%%%%%%%%%%%%%%%%%%%%%%%%%%%%%%%%%%%%%%%%%%
%

%
%%%%%%%%%%%%%%%%%%%%%%%%%%%%%%%%%%%%%%%%%%%%%%%%%%%%%%%%%%%%%%%%%%%%%%
% TABLE 3
%%%%%%%%%%%%%%%%%%%%%%%%%%%%%%%%%%%%%%%%%%%%%%%%%%%%%%%%%%%%%%%%%%%%%%
%

\begin{deluxetable}{llcccccc}
%\rotate
\tablecolumns{8}
\tabletypesize{\scriptsize}
\tablewidth{0pt}
\tablecaption{Deep VRO sample}
\tablehead{
\multicolumn{2}{c}{Optical coordinates} &
\multicolumn{4}{c}{Optical magnitudes} &
\colhead{Photometric} &
\colhead{}\\
\multicolumn{1}{c}{$\alpha_{2000}$}&
\multicolumn{1}{c}{$\delta_{2000}$}&{$I_{F814W}^{\rm a}$}&{$K^{\rm a}$}&{$I-K^{\rm b}$}&{$V_{F606W}^{\rm a}$}&{Redshift$^{\rm c}$}&{Other $\lambda^{\rm d}$}}
\startdata
         12 36 41.39&                   +62 11 43.0$^{\rm f}$&                23.0&                18.9&                 4.1&                24.0&     1.524$^{\rm e}$&                 R,Z\\
         12 36 44.12&                   +62 12 44.8$^{\rm f}$&                24.6&                19.6&                 4.8&                26.6&                1.32&                    \\
         12 36 44.56&                   +62 12 15.4$^{\rm f}$&                25.3&                20.7&                 4.6&                26.7&                1.44&                    \\
         12 36 45.65&                   +62 12 41.9$^{\rm f}$&                25.5&                20.0&                 5.4&                27.0&                1.64&                    \\
         12 36 48.62&                   +62 12 15.7$^{\rm f}$&                24.3&                19.8&                 4.4&                25.4&                1.60&                    \\
\\
         12 36 52.09&                             +62 12 26.2&                23.8&                19.7&                 4.0&                25.9&                1.00&                 X,R\\
         12 36 52.10&                   +62 12 01.1$^{\rm f}$&                27.4&                21.5&                 5.7&                29.1&                1.84&                    \\
         12 36 56.14&                   +62 13 25.1$^{\rm f}$&                24.9&                20.4&                 4.3&                26.9&                1.08&                    \\
         12 36 56.58&                             +62 12 57.3&                25.3&                21.0&                 4.1&                27.4&                1.00&                    \\
\enddata

\tablenotetext{a}{Vega-based magnitude converted from AB magnitude given in Fern\'andez-Soto et~al. (1999).}
\tablenotetext{b}{Calculated $I-K$ color. The $I_{F814W}$ magnitudes have been converted to the $I$ band following Cowie, Songalia, \& Barger (1999); see \S2.}
\tablenotetext{c}{Seven-band photometric redshift from Fern\'andez-Soto et~al. (1999).}
\tablenotetext{d}{Source has multi-wavelength counterpart. (X) X-ray counterpart, see Table~4; (R) Radio counterpart (Richards et~al. 1998; Richards 2000); (Z) Spectroscopic redshift (Dawson et~al. 2001).}
\tablenotetext{e}{Given redshift is spectroscopic.}
\tablenotetext{f}{Source used in stacking analysis; see \S3.3.2.}

\end{deluxetable}

%
%%%%%%%%%%%%%%%%%%%%%%%%%%%%%%%%%%%%%%%%%%%%%%%%%%%%%%%%%%%%%%%%%%%%%%
% TABLE 4: X-ray detected VROs
%%%%%%%%%%%%%%%%%%%%%%%%%%%%%%%%%%%%%%%%%%%%%%%%%%%%%%%%%%%%%%%%%%%%%%
%

%
%%%%%%%%%%%%%%%%%%%%%%%%%%%%%%%%%%%%%%%%%%%%%%%%%%%%%%%%%%%%%%%%%%%%%%
% TABLE 4
%%%%%%%%%%%%%%%%%%%%%%%%%%%%%%%%%%%%%%%%%%%%%%%%%%%%%%%%%%%%%%%%%%%%%%
%

\begin{deluxetable}{llcccccccccccccccc}
\rotate
\tablecolumns{18}
\tabletypesize{\tiny}
\tablewidth{0pt}
\tablecaption{X-ray detected VROs}
\tablehead{
\multicolumn{2}{c}{X-ray coordinates} &
\multicolumn{4}{c}{Optical magnitudes} &
\multicolumn{3}{c}{X-ray counts}      &
\colhead{Band} & 
\colhead{} & 
\multicolumn{3}{c}{X-ray flux}      &
\multicolumn{2}{c}{FB luminosity} &
\colhead{} &
\colhead{VRO} \\ 
\colhead{$\alpha_{2000}$}       &
\colhead{$\delta_{2000}$}       &
\colhead{$I^{\rm a}$}                  &
\colhead{$HK^{\prime \rm a}$}                  &
\colhead{$I-K^{\rm b}$}                 &
\colhead{$V^{\rm a}$}                  &
\colhead{FB$^{\rm c}$}          &
\colhead{SB$^{\rm c}$}          &
\colhead{HB$^{\rm c}$}          &
\colhead{Ratio$^{\rm d}$}       &
\colhead{$\Gamma^{\rm e}$}       &
\colhead{FB$^{\rm f}$}          &
\colhead{SB$^{\rm f}$}          &
\colhead{HB$^{\rm f}$}          &
\colhead{$z=1^{\rm g}$}         &
\colhead{$z=3^{\rm g}$}         &
\colhead{Other $\lambda^{\rm h}$} &
\colhead{Sample$^{\rm i}$}} 
\startdata
              12 36 15.89&                        +62 15 15.5&             $>$25.3&                21.4&              $>$4.2&             $>$26.5&       81.5$\pm$18.7&             $<$20.7&       71.9$\pm$20.7&                       $>$3.52&           $<$--0.26&                2.19&             $<$0.10&                2.48&                2.28&               43.45&                   R&                   I\\
              12 36 20.52&                        +62 12 38.1&                24.7&                20.4&                 4.6&                27.0&        29.4$\pm$7.9&        16.3$\pm$5.8&        15.0$\pm$6.3&        $0.93^{+0.65}_{-0.46}$&                0.88&                0.49&                0.09&                0.44&                0.40&                7.64&                    &                 I,M\\
              12 36 33.76&                        +62 13 13.8&                24.7&                20.7&                 4.3&             $>$26.5&        40.3$\pm$8.3&              $<$9.8&        35.6$\pm$7.9&                       $>$3.67&           $<$--0.29&                1.06&             $<$0.05&                1.19&                1.11&               21.14&                    &                   I\\
              12 36 35.27&                        +62 11 51.7&                24.5&                20.8&                 4.1&                27.4&        23.9$\pm$7.0&         9.9$\pm$4.8&        12.7$\pm$5.8&        $1.29^{+1.37}_{-0.73}$&                1.40&                0.29&                0.05&                0.32&                0.25&                4.79&                    &                   I\\
              12 36 39.65&                        +62 09 36.7&             $>$25.3&                21.1&              $>$4.5&             $>$26.5&        14.3$\pm$6.9&         9.5$\pm$5.3&             $<$12.1&                       $<$1.29&                1.40&                0.19&                0.06&             $<$0.33&                0.17&                3.17&                    &                   I\\
\\
              12 36 42.89&              +62 09 58.9$^{\rm j}$&                25.7&                19.8&                 6.2&             $>$26.5&             $<$11.8&         4.1$\pm$2.0&              $<$9.8&                       $<$2.40&                1.40&             $<$0.16&                0.02&             $<$0.27&             $<$0.15&             $<$2.90&                    &                   M\\
              12 36 47.95&                        +62 10 19.9&             $>$25.3&                21.0&              $>$4.6&             $>$26.5&        36.2$\pm$8.3&        11.6$\pm$5.3&        24.4$\pm$7.2&        $2.11^{+1.87}_{-0.91}$&                0.10&                0.96&                0.07&                0.89&                0.90&               17.20&                   I&                   I\\
              12 36 51.97&              +62 12 25.8$^{\rm j}$&                23.7&      20.0$^{\rm k}$&                 4.0&                26.9&              $<$7.3&         3.6$\pm$2.0&              $<$4.0&                       $<$1.12&                1.40&             $<$0.08&                0.02&             $<$0.09&   $<$0.03$^{\rm l}$&                 $-$&                 R,Z&                   D\\
              12 36 54.57&                        +62 11 11.0&                25.3&                21.4&                 4.2&                26.0&        30.7$\pm$7.7&        12.1$\pm$5.2&        20.6$\pm$6.7&        $1.70^{+1.38}_{-0.75}$&                0.28&                0.77&                0.07&                0.75&                0.70&               13.32&                    &                   I\\
              12 36 58.83&                        +62 10 22.3&                24.1&                20.4&                 4.0&                26.5&        42.6$\pm$8.9&             $<$10.3&        39.9$\pm$8.6&                       $>$3.87&           $<$--0.34&                1.20&             $<$0.05&                1.41&                1.28&               24.39&                    &                 I,M\\
\\
              12 37 01.66&              +62 11 45.9$^{\rm j}$&                24.8&                19.5&                 5.6&             $>$26.5&        11.8$\pm$4.4&         7.2$\pm$2.8&              $<$7.4&                       $<$1.03&                1.40&                0.14&                0.04&             $<$0.18&      0.04$^{\rm l}$&                 $-$&               R,I,Z&                   M\\
              12 37 07.23&                        +62 14 08.0&                25.0&                20.4&                 5.0&                26.3&        54.7$\pm$9.3&        24.2$\pm$6.4&        32.2$\pm$7.7&        $1.35^{+0.58}_{-0.43}$&                0.50&                1.00&                0.12&                0.92&                0.88&               16.75&                   R&                 I,M\\
              12 37 14.07&                        +62 09 16.7&                23.9&                20.2&                 4.0&                25.8&      333.2$\pm$21.2&      223.7$\pm$16.8&      111.2$\pm$13.3&        $0.50^{+0.07}_{-0.07}$&                1.46&                3.67&                1.11&                2.58&                2.44&               46.50&                    &                 I,M\\
\enddata

\tablenotetext{a}{Vega-based magnitude.}
\tablenotetext{b}{Calculated $I-K$ color. The $HK^{\prime}$-band magnitudes are converted to $K$-band magnitudes following Barger et~al. (1999); see Table~1.}
\tablenotetext{c}{Source counts and errors taken from Paper V for those sources detected with a false-positive probability threshold of 10$^{-7}$ and taken from {\sc wavdetect} for those sources detected with a threshold of 10$^{-4}$. ``FB'' indicates full band (0.5--8.0~keV), ``SB'' indicates soft band (0.5--2.0~keV), and ``HB'' indicates hard band (2.0--8.0~keV).}
\tablenotetext{d}{Ratio of the count rates between the 2.0--8.0~keV and 0.5--2.0~keV bands. The errors were calculated following the ``numerical method" described in \S1.7.3 of Lyons (1991).}
\tablenotetext{e}{Photon index for the 0.5--8.0~keV band, calculated from the band ratio. Taken from Paper V for those sources detected with a false-positive probability threshold of 10$^{-7}$. The photon indices for the X-ray sources detected with a threshold of 10$^{-4}$ and those with a low number of counts have been set to $\Gamma =1.4$, a value representative of the X-ray background spectral slope; see Paper V.}
\tablenotetext{f}{Fluxes are in units of $10^{-15}$~erg~cm$^{-2}$~s$^{-1}$. These fluxes have been taken from Paper V and they have not been corrected for Galactic absorption.}
\tablenotetext{g}{Predicted rest-frame unabsorbed 0.5--8.0~keV luminosities in units of $10^{43}$~erg~s$^{-1}$ at $z=1$ and $z=3$. The effect of absorption has been corrected for using the technique described in \S3.2.1.}
\tablenotetext{h}{Source has multi-wavelength counterpart. (R) Radio counterpart (Richards et~al. 1998; Richards 2000); (I) Mid-infrared counterpart (Aussel et~al. 1999a); (Z) Spectroscopic or photometric redshift (Fern\'andez-Soto et~al. 1999; Cohen et~al. 2000).}
\tablenotetext{i}{VRO sample from which the source has been taken. ``M'' indicates moderate-depth sample (see Table~2), ``D'' indicates deep sample (see Table~3), and ``I'' indicates intermediate-depth X-ray detected VRO sample.}
\tablenotetext{j}{X-ray counterpart detected with {\sc wavdetect} with false-positive  probability threshold of 10$^{-4}$. When a source is not detected in a given band an upper limit is calculated following \S3.2.1 in Paper V.}
\tablenotetext{k}{The $HK^{\prime}$ magnitude has been converted from the $K$-band following Barger et~al. (1999); see Table~3.}
\tablenotetext{l}{The rest-frame 0.5--8.0~keV luminosity is given at the source redshift and is calculated assuming $\Gamma=2.0$ and no intrinsic absorption; see \S3.2.2.}

\end{deluxetable}

%
%%%%%%%%%%%%%%%%%%%%%%%%%%%%%%%%%%%%%%%%%%%%%%%%%%%%%%%%%%%%%%%%%%%%%%
% TABLE 5: Stacking results
%%%%%%%%%%%%%%%%%%%%%%%%%%%%%%%%%%%%%%%%%%%%%%%%%%%%%%%%%%%%%%%%%%%%%%
%

%%%%%%%%%%%%%%%%%%%%%%%%%%%%%%%%%%%%%%%%%%%%%%%%%%%%%%%%%%%%%%%%%%%%%%
% TABLE 5
%
% Results of 970 ks stacking analysis...
%%%%%%%%%%%%%%%%%%%%%%%%%%%%%%%%%%%%%%%%%%%%%%%%%%%%%%%%%%%%%%%%%%%%%%
%

\begin{deluxetable}{lcccccccccccrrr}
\rotate
\tabletypesize{\scriptsize}
\tablewidth{0pt}
\tablecaption{X-ray stacking-analysis results for individually undetected VROs}

\tablehead{
\colhead{VRO}&
\colhead{} &
\colhead{Effective}&
\multicolumn{3}{c}{Source counts}&
\multicolumn{3}{c}{Background counts}&
\multicolumn{3}{c}{Statistical probability}&
\multicolumn{3}{c}{Flux}\\
\colhead{Sample} &
\colhead{$N^{\rm a}$}& 
\colhead{Exposure$^{\rm b}$}& 
\colhead{FB$^{\rm c}$}& 
\colhead{SB$^{\rm c}$}&  
\colhead{HB$^{\rm c}$}&  
\colhead{FB$^{\rm d}$}& 
\colhead{SB$^{\rm d}$}&  
\colhead{HB$^{\rm d}$}&  
\colhead{FB$^{\rm e}$}& 
\colhead{SB$^{\rm e}$}&  
\colhead{HB$^{\rm e}$}&  
\colhead{FB$^{\rm f}$}& 
\colhead{SB$^{\rm f}$}&  
\colhead{HB$^{\rm f}$}
}
\startdata
Moderate-depth                      & 18 & 15.8 & 124  & 38 &  67 & 94.3 & 19.6 & 55.2 & $2.2\times10^{-3}$ & $2.0\times10^{-4}$ & 0.07 & 1.5 & 0.6 & $<$2.6 \\
Moderate-depth ($I\le24.1$)         & 9  &  8.1 &  71  & 23 &  39 & 47.7 &  9.3 & 28.6 & $4.0\times10^{-4}$ & $1.0\times10^{-4}$ & 0.04 & 2.2 & 0.9 & $<$3.7 \\
Moderate-depth ($I>24.1$)           & 9  &  7.7 &  53  & 15 &  28 & 46.5 &  10.3 & 26.5 & 0.20 & 0.11 & 0.41 & $<$2.0 & $<$0.6 & $<$3.7 \\
Deep                                & 7  &  6.5 &  52  & 16 &  23 & 38.4 &  7.7 & 22.4 & 0.02 & $5.0\times10^{-3}$ & 0.48 & $<$2.1 & 0.7 & $<$4.0 \\
\enddata
\tablenotetext{a}{Total number of sources used in the stacking analysis.}
\tablenotetext{b}{Effective \chandra\ exposure time in Ms.}
\tablenotetext{c}{Total counts measured. ``FB'' indicates full band, ``HB'' indicates hard band, and ``SB'' indicates soft band.}
\tablenotetext{d}{Average number of background counts found using the Monte-Carlo simulations described in \S3.3. A local background is used as per \S2 of Brandt et~al. (2001c, hereafter Paper VII).}
\tablenotetext{e}{Monte-Carlo probability that the total number of counts could be due to statistical chance.}
\tablenotetext{f}{Average flux or $3\sigma$ upper limit in units of $10^{-17}$~erg~cm$^{-2}$~s$^{-1}$; all fluxes calculated assuming $\Gamma=2.0$.}

% \newpage

\end{deluxetable}

%%%%%%%%%%%%%%%%%%%%%%%%%%%%%%%%%%%%%%%%%%%%%%%%%%%%%%%%%%%%%%%%%%%%%%

\end{document}